\documentclass[conference]{IEEEtran}
\IEEEoverridecommandlockouts
\usepackage{cite}
\usepackage{amsmath,amssymb,amsfonts}
\usepackage{algorithmic}
\usepackage{graphicx}
\usepackage{textcomp}
\usepackage{xcolor}
\usepackage{amsfonts,amssymb} %空心包
\usepackage{bbold}%空心包
\usepackage{amsmath,amssymb} %集合包
\usepackage{amsthm,amsmath,amssymb}%花写字母
\usepackage{mathrsfs} %花写字母
\usepackage{threeparttable} %表格注释
\usepackage{float}
\usepackage{booktabs}
\usepackage{multirow}%合并单元格
\usepackage{bbding}%打勾
\usepackage{stfloats}%跨栏
\usepackage{svg}
\usepackage[colorlinks,
            linkcolor=blue,
            anchorcolor=blue,
            citecolor=blue]{hyperref}
\usepackage{algorithm}
\usepackage{algorithmic}

\def\BibTeX{{\rm B\kern-.05em{\sc i\kern-.025em b}\kern-.08em
    T\kern-.1667em\lower.7ex\hbox{E}\kern-.125emX}}
\begin{document}

%标题
%\title{Group Meiosis Contrast Model: Self-supervised Learning For EEG-Based Emotion Recognition\\
\title{Self-supervised Group Meiosis Contrastive Learning for EEG-Based Emotion Recognition\\
\thanks{*Corresponding author\\$^1$ https://github.com/kanhaoning/Self-supervised-group-meiosis-contrastive-learning-for-EEG-based-emotion-recognition}
}

%\author{\IEEEauthorblockN{1\textsuperscript{st} Haoning Kan }
\author{\IEEEauthorblockN{ Haoning Kan }
\IEEEauthorblockA{\textit{Faculty of Science} \\
%\IEEEauthorblockA{\textit{Haoning Kan} \\
%\textit{Beijing University of Technology}\\
\textit{Beijing University of Technology}\\
Beijing, China \\
KanHaoning@emails.bjut.edu.cn}
\and
\IEEEauthorblockN{ Jiale Yu}
\IEEEauthorblockA{\textit{Beijing-Dublin International College} \\
%\IEEEauthorblockA{\textit{} \\
\textit{Beijing University of Technology}\\
Beijing, China \\
jiale.yu@ucdconnect.ie}
\and
\IEEEauthorblockN{ Jiajin Huang}
\IEEEauthorblockA{\textit{Faculty of Information Technology} \\
\textit{Beijing University of Technology}\\
Beijing, China \\
jhuang@bjut.edu.cn}
\and
\IEEEauthorblockN{ Zihe Liu }
\IEEEauthorblockA{\textit{Faculty of Science} \\
\textit{Beijing University of Technology}\\
Beijing, China \\
Zihe\_Liu@emails.bjut.edu.cn}
\and
\IEEEauthorblockN{ Haiyan Zhou* }
\IEEEauthorblockA{\textit{Faculty of Information Technology} \\
\textit{Beijing University of Technology}\\
Beijing, China \\
zhouhaiyan@bjut.edu.cn}

}

\maketitle

\begin{abstract}
\par%背景、意义解释清楚型
\par%简化背景意义强调工作重点型
The progress of EEG-based emotion recognition has received widespread attention from the fields of human-machine interactions and cognitive science in recent years. However, how to recognize emotions with limited labels has become a new research and application bottleneck. To address the issue, this paper proposes a Self-supervised Group Meiosis Contrastive learning framework (SGMC) based on the stimuli consistent EEG signals in human being. In the SGMC, a novel genetics-inspired data augmentation method, named Meiosis, is developed. It takes advantage of the alignment of stimuli among the EEG samples in a group for generating augmented groups by pairing, cross exchanging, and separating. And the model adopts a group projector
%is designed 
to extract group-level feature representations from group EEG samples triggered by the same emotion video stimuli. 
Then contrastive learning is employed to maximize the similarity of group-level representations of augmented groups with the same stimuli.
%Then positive pairs are formed with augmented groups with the same stimuli. Contrastive learning is further employed to maximize the similarity of group-level representations between positive pairs. 
%augmented groups with the same stimuli. 
The SGMC achieves the state-of-the-art emotion recognition results on the publicly available DEAP dataset with an accuracy of 94.72\% and 95.68\% in valence and arousal dimensions, and also reaches competitive performance on the public SEED dataset with an accuracy of 94.04\%. 
%The SGMC reaches competitive emotion recognition performance on publicly available DEAP and SEED datasets.
It is worthy of noting that the SGMC shows significant performance even when using limited labels. Moreover, the results of feature visualization suggest that the model might have learned video-level emotion-related feature representations to improve emotion recognition. 
And the effects of group size are further evaluated in the hyper parametric analysis.
%Further, the hyper parametric analysis evaluates the meaningful law of combination between group size and other critical hyper parameters.  %Further, we explored the reason for significant results through feature visualization and hyper parametric analysis. 
Finally, a control experiment and ablation study are carried out to examine the rationality of architecture. 
%Finally, the control experiment and the ablation study verify the rationality of architecture. 
The code is provided publicly online$^1$.

%A group-projector-based model is adopted. The model achieves extracting a group-level representation by extracting individual representations from a group and aggregating them into a group-level representation. % and organically combined it with self-supervised contrastive learning.
\par
\end{abstract} 
%关键词
%引言
\section{Introduction}
\par
%Emotion
\par
	Emotion plays a crucial role in human cognition  %in rational decision-making, perception, interpersonal communication, and human intelligence. 
% and provides guides 
and involves many application fields. For example, in the field of human-machine interaction \cite{b55}, emotion recognition enables the machine to provide more humanized interaction. In consumer neuroscience, emotion analysis is a common tool to obtain the user experience for product design\cite{b54}. %Therefore, understanding emotion is of great significance for research on cutting-edge technology and the decision-making of traditional commerce.
 Recently, the method of emotion recognition based on Electroencephalography (EEG) signal has shown its advantages. Compared to conscious behavior signals such as facial expression and body language, the EEG has the advantage of being difficult to hide or disguise. Compared with other physiological signals such as the fMRI (functional magnetic resonance imaging), and ECG (Electrocardiogram), the EEG is more convenient for sampling and has a higher time resolution. 
\par
There is great progress in the field of EEG-based emotion recognition. With traditional machine learning techniques, the handcrafted features are calculated and selected carefully, which is quite critical during emotion recognition.
%There are many EEG-based emotion recognition methods. 
%Based on the handcrafted features, there is great progress in EEG-based emotion recognition with traditional machine-learning methods. 
While these approaches relie too much on the researcher's experiences on EEG signals and cognitive related knowledge.
%The traditional machine-learning methods rely on the handcrafted feature to classify would obtain a satisfactory result, but also rely on expert experience. 
In recent years, the development of deep learning methods achieves competitive accuracy, which could not pay attention to the handcrafted features.
%In recent years, the development of deep learning methods contributes to many deep learning models to achieve competitive accuracy on EEG-based emotion recognition without complex handcrafted features.
With the guidance of a large number of data with labels, the deep learning models would learn high-level emotion-related feature representation for precise affective computing \cite{b12,b13,b14,b15,b16}.
\par
Generally, artificial labels are crucial for training deep learning models based on the common supervised methods. 
But there is some condition requiring higher accurate and real-time recognition, 
%But in some fields requiring higher granularity and real-time, 
obtaining qualified labels is expensive. %It results that labels which can ensure the training of models which meet requirements are scarce.
For example, in neuroscience, EEG is frequently used to explore the process of emotion, such as in the tasks of empathy and reading comprehension. Participants are usually required to answer questions, so that the researchers could get their emotion situations. However, the emotion labels obtained through this way are time-consuming and laborious. And it is easy to generate subjective bias, which might decrease the reliability of labels \cite{b5,b6}.
%For example, in some research in cognitive science, EEG is used to explore more granular emotions and cognitive activities, such as empathy and reading comprehension. 
%In such kinds of experiments, subjects are usually required 
%to fill out complex questionnaires, which makes labeling costly and increases subjectivity decreases the qualities of labels \cite{b5}\cite{b6}. 
%so that the quality of the label may decline. \cite{b5}\cite{b6}. %The high cost makes labeled data scarce. On the contrary, the cost of recording unlabeled EEG data is much lower. 
Similarly, in the application of consumer neuroscience, the EEG signals are recorded to evaluate the participants' emotional states while they are playing games, listening to music, and watching movies, and advertisements,  which aims to provide instructive references to the content creator, \cite{b1,b2,b3,b4}.
%Similarly, in the application of consumer neuroscience, the EEG signals are applied to probe the experiencer's emotional state in real-time while playing games, listening to music, and watching movies, and advertisements, which aims to provide instructive references to the content creator, \cite{b1}-\cite{b4}. 
%It requires more accurate and intensive emotional labels. %However such labels are costly to obtain and often scarce.
In these conditions, the precise and time-intensive labels are also required.
Therefore, the lack of qualified labels hinders the application of machine learning-based models in many precise fields. 
%\par
\par
Previous studies have explored to reduce the dependence on artificial labels \cite{b7,b8}. Several neuroscience studies have shown the exploitable consistency of stimuli in emotion EEG signals. 
% is exploitable to make up for the lack of artificial labels. 
They have discovered that the EEG signals among a group of subjects who watched the same emotional video clips share similar group-level stimuli-related features \cite{b46,b47}. Such features correlated with preference, arousal, valence, etc, are potential to make up for the lack of artificial labels.
%Previous studies have explored to reduce the dependence on artificial labels \cite{b7}\cite{b8}. Several pioneering neuroscience studies have shown the nature existing in most emotion EEG datasets has the potential to transform into potential labels to make up for the lack of artificial labels. \cite{b46}\cite{b47} discovered that EEG signals among a group of subjects who watched the same emotional video clips share similar group-level stimuli-related features. This indicates such a group of EEG signals that share the same stimuli label is correlated with preference, arousal, valence, etc. Although the inherent nature of such stimuli labels endows itself with higher granularity and intensity than artificial labels but also make it more difficult to directly utilize for improving emotion recognition.
	 Existing methods mainly adopt the self-supervised learning (SSL) method to exploit such stimuli consistency. SSL can generate labels according to the attributes of data for learning. For example, shen et al. proposed a novel contrastive learning framework \cite{b9} to learn representation by making the model maximizing the similarity between representations of EEG signals corresponding to the same stimuli. %, and %minimum the similarity
%minimizing the similarity of representations corresponding to different stimuli. 
However, there exist random effects in the emotion-related EEG signals. For example, whether the subjects were distracted during the emotional tasks and their fatigue situations would increase the noise of the signals. And also the responses of participants could not be totally the same, which increases the difficulty in maximizing the similarity across subjects in contrastive learning.
 
%However, some obstacles that hinder the usual contrastive learning framework training. There are individual differences between subjects, so the model is hard to predict whether two signals correspond to the same stimuli in the SSL training.  Moreover, random distraction and fatigue further increase the difficulty of contrastive learning.
\par
%This problem results in strict requirements for SSL training. For example, it needs to eliminate some samples with weak emotions, which leads to reducing the data utilization. To tackle these obstacles, 
To further improve the EEG-based emotion recognition under the SSL framework,
we proposed a Self-supervised Group Meiosis Contrastive learning (SGMC) framework for EEG-based emotion recognition.

First, since larger samples could be better to represent the characteristics of signals from the view of statistics, we design a group projector in SGMC to collect a group of EEG samples to extract group-level representation for contrastive learning.
 
%Utilizing group-level representation to represent a group of EEG samples corresponding to the same stimuli label could mitigate such hinders (individual difference, distraction, and so on). Because although the individual differences are significant, the statistical features of the subject group are more stable. So we design a novel group projector to aggregate such a sample group to extract group-level representation for contrastive learning. 
\par
	Second, we proposed a novel method of data augmentation to provides augmented group samples for contrastive learning. Applying data augmentation to enhance contrastive learning is a basic paradigm, however, there are few studies on augmenting group samples.
Inspired by the meiosis mechanism in genetics \cite{b11}, we augment data without changing stimuli features by pairing, cross exchanging, and separating. In this way, data augmentation enables contrastive learning further take the advantage of the alignment of stimuli in the EEG signal group. And then the SGMC enables the model learn critical representations and achieve competitive emotion recognition performance with a significant improvement. 
%Genetics provides a novel inspiration for us. Meiosis is the key mechanism in genetics to provide diverse gene recombination and is the driving force to increase species diversity\cite{b11}. In the proposed contrastive learning process, two EEG signals of the positive pair are aligned in the video time which is similar to the alignment of two alleles in the space of two homologous chromosomes. So we model the meiosis of chromosomes to design a data augmentation method for contrastive learning. In this way, the Meiosis data augmentation makes full use of the advantage of the alignment of stimuli among the group sample to improve contrastive learning and enable the model to achieve excellent performance on emotion recognition. 
Here we summarize the contributions of this paper as follows:
\par
\begin{itemize}
\item To reduce the dependence on emotion labels, we introduce a self-supervised contrastive learning framework to further exploit the consistency of stimuli for EEG-based emotion recognition. %On the SEED dataset, based on pre-training, the model achieves 91.01\% accuracy fine-tuned with 50 labeled samples per category (0.14\% of the full training set), exceeding 89.83\% of fully-supervised learning with the full training set.
\item To decrease the effects of individual difference and random effects in EEG signals, we design a group-based contrastive learning framework to extract group-level stimuli-related feature representations. 
\item To augment the group sample, we design a genetics-inspired data augmentation method, named Meiosis. It utilizes the alignment of stimuli to augment group samples without changing stimuli features. which provides augmented group samples to enhance contrastive learning.
%\item We design a group contrastive learning framework for extracting group-level stimuli-related feature representations from group EEG samples. It alleviates the problem that the stimuli-related is difficult to extract from the individual sample for contrastive learning.
%\item  We design a novel genetics mechanism inspired data augmentation Meiosis to take advantage of the alignment of stimuli among the group sample, and organically combine it with self-supervised contrastive learning. It significantly improves the performance of emotion recognition. 
%\item The experiments demonstrate that SGMC achieves excellent performance in limited labeled sample learning. On the SEED dataset, based on exploiting potential stimuli labels to pre-training, the model achieves 91.01\% accuracy fine-tuned with 50 artificial labeled samples per category (0.14\% of the full training set), exceeding 89.83\% of fully-supervised learning with the full training set.
\item The SGMC achieves the state-of-the-art emotion classification results on the publicly available DEAP dataset with an accuracy of 94.72\% and 95.68\% in valence and arousal dimensions. On public SEED dataset also reaches competitive 94.04\% accuracy, and achieve 91.01\% accuracy fine-tuned with 50 labeled samples per category (0.14\% of the full training set), exceeding 89.83\% accuracy of fully-supervised learning with the full training set.
\end{itemize}
%相关工作
\section{Related Work}
\subsection{EEG-based Emotion Recognition}
\par
	In earlier studies, emotional features of EEG signals were usually extracted to recognize by some traditional machine learning strategies. Such as the support vector machine (SVM)\cite{b48}, Gaussian Naive Bayes classification \cite{b41}, and k-nearest neighbor (k-NN)\cite{b49} are widely used classify emotion of the EEG signal. 
\par
	Compared with the traditional machine learning method, the deep learning model has more advantages in extracting high-level emotional features. In recent years, more and more deep learning neural networks based on emotion recognition models achieved good performance on EEG-based emotion recognition tasks \cite{b12,b13,b14,b15,b16}. 
\par
	Recently popular methods focus on recurrent neural networks (RNNs/LSTMs), and convolutional neural networks (CNNs). In 2017, Alhagry et al \cite{b53} adopted a two-layer long-short term memory (LSTM) to reach satisfactory emotion classification with the input of the raw EEG signals. In 2020, Li et al \cite{b39} constructed model BiHDM adopted four RNN modules to capture the input of each hemispheric EEG electrode's data from horizontal and vertical streams and achieved the SOTA. CNN is also widely used for extracting spatial features of the EEG signal. In 2016, Li et al \cite{b18} proposed a hybrid network structure based on CNN and RNN for emotion recognition based on multi-channel EEG signals, which shown the effectiveness of a hybrid network in the trial-level emotion recognition tasks. In 2017, Alhagry et al. \cite{b50} explored a convolutional neural network and a simple deep neural network. This CNN model shown more significant performance and achieved the SOTA. In 2018 Shawky et al. \cite{b19} proposed a 3D CNNs model, which divides raw signals into 6-s segments to input. In the same year, Yang et al. \cite{b28} proposed a hybrid model combining CNN and RNN networks to learn spatial-temporal representation for emotion recognition. It utilized a sparse matrix as input to reflect the relative position of the electrodes. Compared with complex input of RNNs and 2D/3D CNNs, Cheah et al. \cite{b20} proposed a 1D-CNN based ResNet18, which adopted simple input(channel $\times$ time) to train the deeper neural network. It is more suitable to perform pre-training with simple data processing and a faster training process.
\subsection{Self-supervised Learning}
\par
	Self-supervised learning aims to learn representation without relying on artificial labels.  
The latest research in the field of machine learning and deep learning shown the potential of the SSL method in learning generalized and robust representations \cite{b21,b22,b23,b24,b25,b26}. 
	SSL has been widely used in many fields. For example, in computer vision (CV), Gidaris et al. \cite{b21} based on spatial properties designed an SSL task to rotate the original image and require the model to predict the rotation angle. Based on the temporal properties of the video, an SSL task \cite{b22} was designed to require the model to predict whether the two video frames are close in time. 
 In natural language processing (NLP), $word2vec$ \cite{b23} designed SSL tasks such as predicting headword and adjacency words, etc. $BERT$ \cite{b24} designed two SSL tasks masked language prediction and next sentence prediction, and achieved SOTA on 11 NLP tasks. In EEG signal processing, Zhang et al. \cite{b35} applied Generative Adversarial Network to design the SSL method. 
It makes the generator augment masked original signals to get simulated signals and requires the model to distinguish real and simulated signals, which alleviates the problem of EEG data scarcity and achieves SOTA.
\par
Recently contrastive-learning-based SSL has made progress in EEG signal processing. 
Contrastive learning defines any two samples with internal relations as the positive pair, otherwise, it is the negative pair, whose loss function aims to maximize the similarity of representations between positive pairs minimums the similarity of representation between negative pairs. Shen et al. \cite{b9} proposed a self-supervised contrastive learning framework CLISA to improve inter-subjects prediction, which requires the model to predict whether two EEG signals are recorded when two subjects watch the same video clip. In this way, the model learned well inter-subject representation ability and achieved SOTA in inter-subject prediction after fine-tuning. In \cite{b7} several self-supervised contrastive learning methods were proposed to improve performance on limited label sample tasks. Among them,  Relative Positioning (RP) requires the model to predict whether the two EEG signals are recorded in close time, and Contrastive Predictive Coding (CPC) requires the model to predict the representation of adjacent EEG signals via the anchor signal. 
They confirmed that models learned physiologically and clinically meaningful feature representations by SSL pre-training without label guidance. Further, they fine-tuned the pre-trained model to significantly outperformed the fully-supervised baseline on less labeled sample learning tasks.
In \cite{b8} an augment-based SSL method is proposed, which requires the model to predict whether two augmented EEG signals come from the same original signal. It applies classical data augmentation such as time warping permutation and crop$\&$resize and so on. The generalization ability of the model has significantly improved and exceeded fully-supervised learning in both the full and the limited labeled sample learning on sleep staging.
Contrastive learning shows its excellence in improving inter-subject prediction, learning physiological feature representation without labels, and so on in EEG signal processing.
%方法	
\\

\section{Proposed Method}
\subsection{Overall Framework}\label{AA}
\par
This paper designs a Self-supervised Group Meiosis Contrastive learning (SGMC) framework for EEG-based emotion recognition. As illustrated in Fig.1 the proposed framework consists of a contrastive learning pre-training process and an emotion recognition fine-tuning process. 
In the pre-training process, SGMC contains five components: a group sampler, the Meiosis data augmentation, a base encoder, a group projector, and a contrastive loss function. Firstly, the group sampler generates a minibatch containing several groups of EEG signals for augmenting. Secondly, the Meiosis augments each EEG group to generate two groups for constructing the positive and negative pairs. Nextly the base encoder extracts individual-level stimuli-related representations from each EEG signal. Then the group projector aggregates each group of
\begin{figure}%[H]
\centering
\includegraphics[scale=0.42]{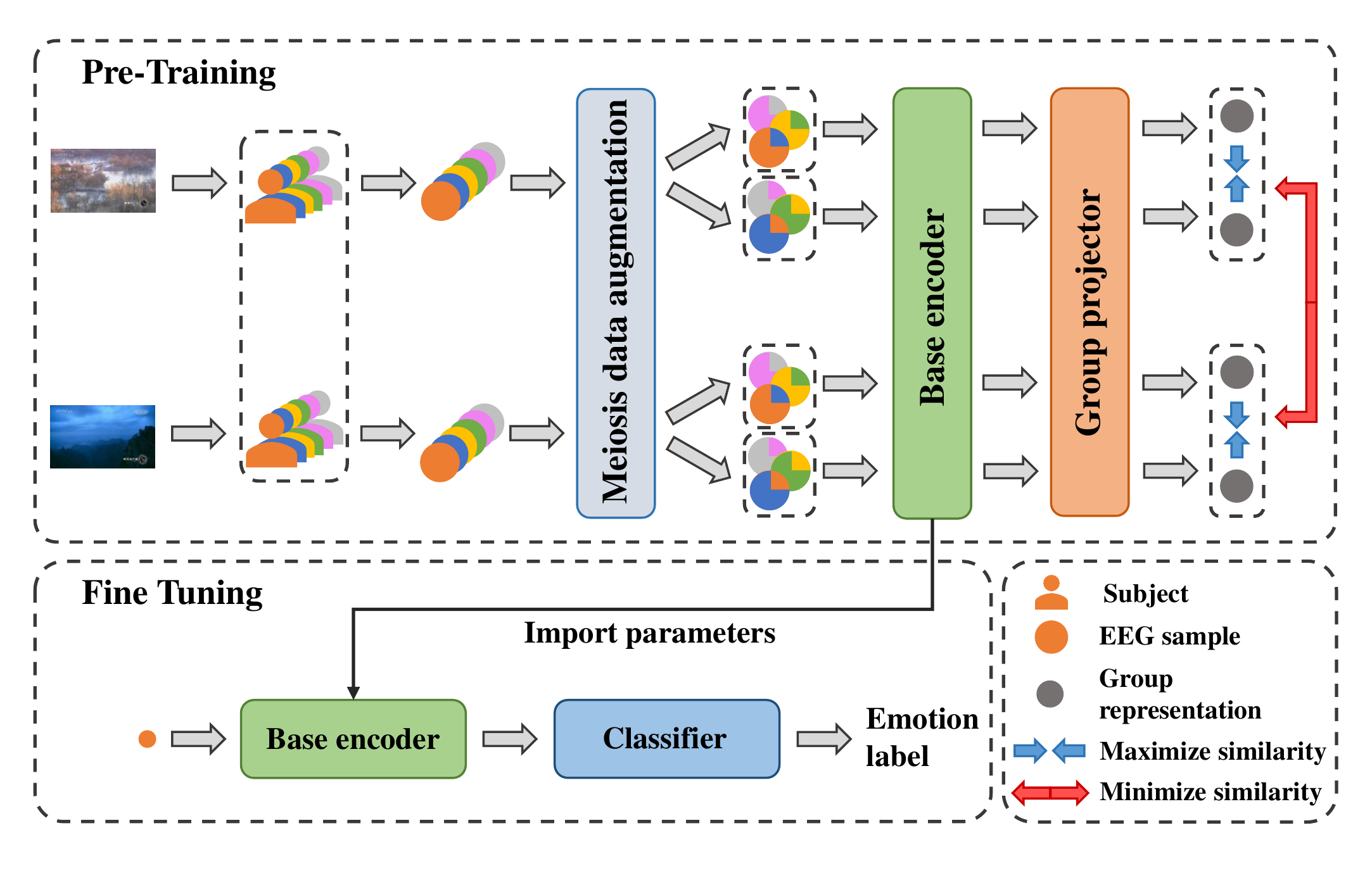}
\caption{Illustration of the proposed SGMC. During the process of pre-training, each group of samples is sampled from EEG samples corresponding to the same video clip stimuli. Then each group of EEG samples is augmented by genetics inspired Meiosis data augmentation to generate two augmented group samples. Each augmented group is sent to the base encoder to extract individual representations of each individual sample and then the group projector aggregates them to obtain the group-level representation. The model is required to maximize the similarity of representations between groups sharing the same stimuli and minimize the representations of groups that correspond to the different stimuli for minimizing the contrastive loss. The pre-trained base encoder will be fine-tuned with a classifier for emotion recognition.}
\label{fig_framework}
\end{figure}
\noindent
 representations to extract group-level stimuli-related representations and map them into another latent space for computing the similarity. Together, the parameters of the base encoder and group projector are optimized by minimizing the contrastive loss. In the fine-tuning process, the model that consisted of the pre-trained base encoder and initialized classifier performs the emotion classification training.

\subsection{Group Sampler}\label{AA}
\par
	Generally, it is difficult to contrastive learning through extracting stimuli-related features from individual EEG samples. So we take the strategy of extracting from group EEG samples, to achieve it we construct the sampler to provide input for the minibatch. 
\par 
In the processed dataset, video clips and subjects correspond to two axes of the dataset tensor. Among it, each EEG sample was defined as $\boldsymbol{X}^{s}_{v}$ $\in$ $\mathbb{R^{M \times C}}$, corresponding to a 1-second signal recorded when subject $s$ watched a 1-second video clip $v$, where $M$ is the number of times samples and $C$ is the dimension of signals ($\it e.g.$,channels). 
To obtain a minibatch, illustrated in Fig.2 sampler first randomly sample $P$ video clips $v_1, v_2, ..., v_P$ that have not been sampled in the current epoch. To sample two equal sample groups to construct positive pair for each clip stimuli, sampler nextly randomly select $2Q$ subjects $s_1, s_2, ..., s_{2Q}$ to prepare for grouping. Further sampler extract the EEG signals corresponding to selected subjects and video clips, $2PQ$ samples $\mathcal{D}=\{\boldsymbol{{X}}^{s_{k}}_{v_{i}}|i=1,2,...,P; k=1,2,...,2Q\}$ are obtained, which were recorded by $2Q$ subjects when watched $P$ video clips respectively. Furthermore, we note a group samples 
$\boldsymbol{G}_{i} = \{\boldsymbol{X}^{s_{1}}_{v_{i}},\boldsymbol{X}^{s_{2}}_{v_{i}},...,\boldsymbol{X}^{s_{2Q}}_{v_{i}}\}$ corresponding to the video clip $v_i$.  Among $\boldsymbol{G}_{i}$, each individual sample shared the similar 
-related features. In this way, sampler would provide the minibatch with $P$ group samples $\{\boldsymbol{G}_{1}, \boldsymbol{G}_{2}, ...,\boldsymbol{G}_{P}\}$ corresponding to $P$ different stimuli for pre-training.
\begin{figure}[H]
\centering
\includegraphics[width=0.48\textwidth,height=0.28\textwidth]{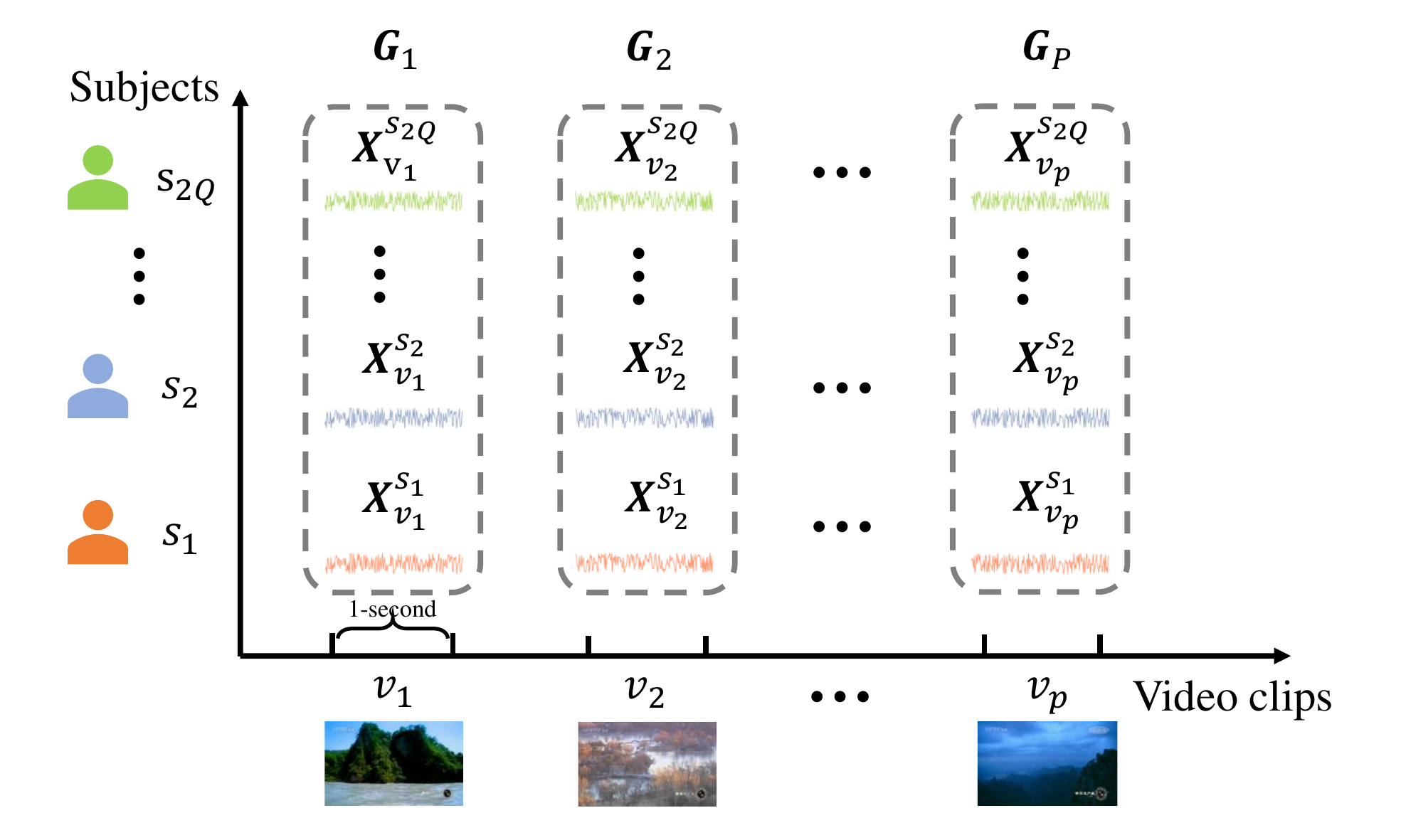}
    \caption{The illustration of sampling for a minibatch. Sampler first samples $P$ video clip and $2Q$ subjects. For each sampled video clip, next the sampler samples a group of EEG signals recorded when sampled $2Q$ subject watched it. Then $P$ groups of EEG samples are obtained for a minibatch.}
\end{figure}
\subsection{Meiosis Data Augmentation}\label{AA} 
\par
%Meiosis aims to take the advantage of the alignment of stimuli among the group sample to improve contrastive learning. It is designed
Meiosis aims to augment one group sample to generate two groups that preserve the same stimuli-related features by utilizing the alignment of stimuli in the EEG group for constructing the positive pair.  
\par
\begin{figure}%[]
\centering
\includegraphics[scale=0.22]{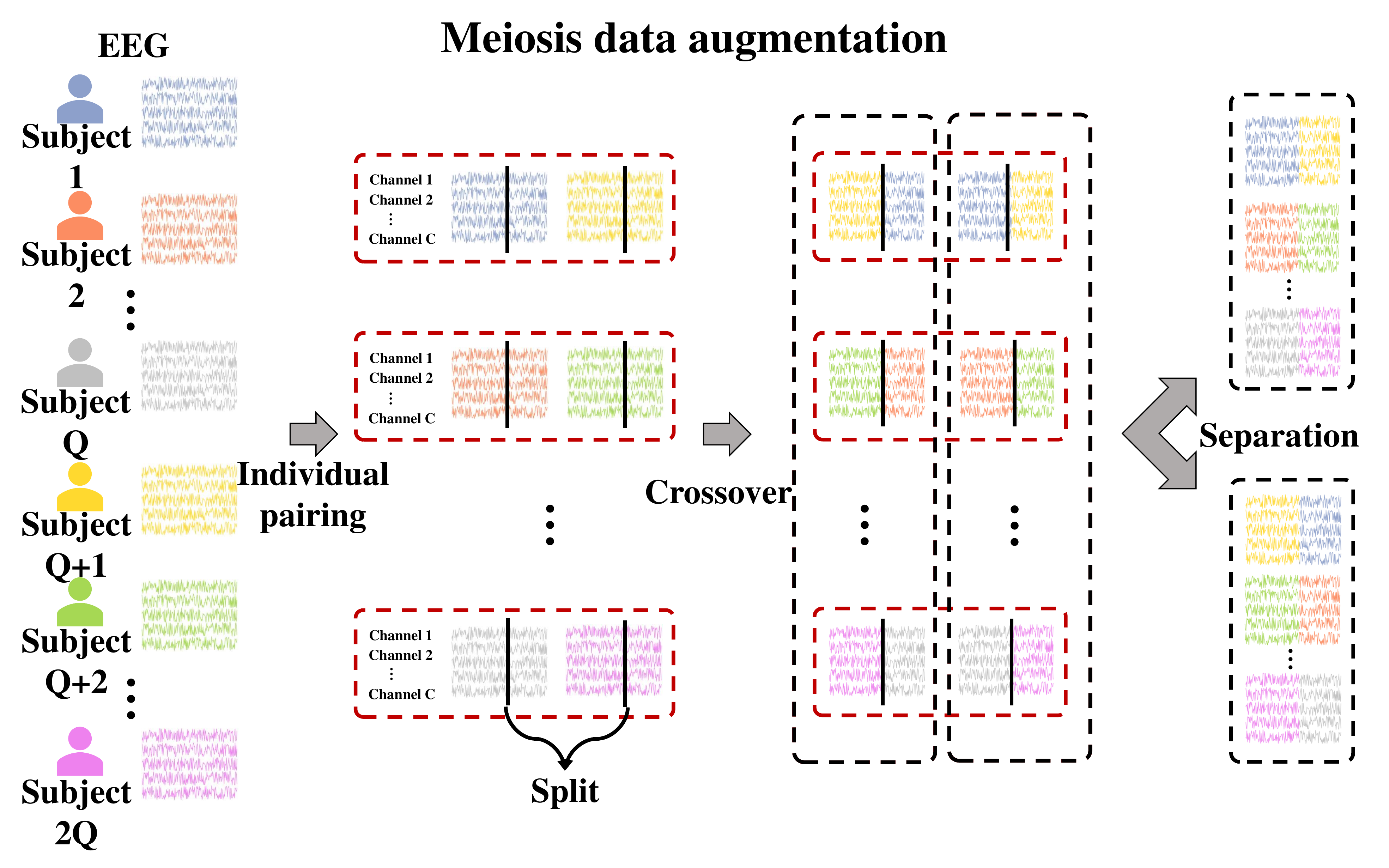}
\caption{The illustration of the Meiosis data augmentation. A group of EEG samples sharing the same stimuli are randomly paired and cross exchanged a part of the signal in a pair, and then separated into two groups.}
\label{fig_framework}
\end{figure}
To increase the meaningful difficulty of the model decoding EEG samples, we hope to mix signals of different subjects. Moreover, to preserve the original stimuli-related features for extraction by SGMC, we select the signals corresponding to the same stimuli to split and splice. So we design the crossover transformation as follows:

\par 
We represent $\{\boldsymbol{a}_1,\boldsymbol{a}_2,...,\boldsymbol{a}_M\}$ as any EEG signal $\boldsymbol{A}$, where $\boldsymbol{a}_i $ is the data at $i^{th}$ sampling point (i=1,2,...,M). Similarly represent $\{\boldsymbol{b}_1,\boldsymbol{b}_2,...,\boldsymbol{b}_M\}$ as any other signal $\boldsymbol{B}$. 
Further we exchange the data of the first $c$ sampling points of two samples $\boldsymbol{A}$ and $\boldsymbol{B}$ to obtain $\widetilde{\boldsymbol{A}}=\{\boldsymbol{b}_1, \boldsymbol{b}_2,..., \boldsymbol{b}_c, \boldsymbol{a}_{c+1}, \boldsymbol{a}_{c+2},...,\boldsymbol{a}_M\}$ and $\widetilde{\boldsymbol{B}}=\{\boldsymbol{a}_1, \boldsymbol{a}_2,...,\boldsymbol{a}_c,\boldsymbol{b}_{c+1}, \boldsymbol{b}_{c+2},..., \boldsymbol{b}_M\}$ , where $c$ is given. Such transformation for any two EEG signals is encapsulated as the following function expression:
\begin{align}
\{\widetilde{\boldsymbol{A}}, \widetilde{\boldsymbol{B}}\}=T({\boldsymbol{A}}, {\boldsymbol{B}}, c)
\end{align}
\par
Furthermore, to take the advantage of the diversity of group combinations, we can randomly pair for crossover and separating. As illustrated in Fig.3  the overall Meiosis data augmentation can be designed as follows:\\
1) Individual pairing: For one original EEG signals group $\boldsymbol{G}_{i}$=$\{\boldsymbol{X}^{s_{1}}_{v_{i}}|k\!=\!1,2,...,2Q\}$ (corresponding to a video clip $v_i$)
 individual signals are randomly paired to form $Q$ pairs 
%, we perform randomly pair the individual signals to obtain $Q$ pairs 
$\{\boldsymbol{X}^{s_1}_{v_i},\boldsymbol{X}^{s_{1+Q}}_{v_i}\},\{\boldsymbol{X}^{s_2}_{v_i},\boldsymbol{X}^{s_{2+Q}}_{v_i}\},...,\{\boldsymbol{X}^{s_Q}_{v_i},\boldsymbol{X}^{s_{2Q}}_{v_i}\}$ for crossover.
\\
2) Crossover : Meiosis receives a randomly given split position $c$ to perform transformation (1) 
for each pairs to obtain \{$\{\widetilde{\boldsymbol{X}}^{s_k}_{v_i}, \widetilde{\boldsymbol{X}}^{s_{k+Q}}_{v_i}\}|k=1,2,...,Q$\}. 
\\
3) Separation: The transformed signals are randomly divided into two groups, and paired transformed signals are required enter into the different groups $A$ and $B$. %Finally we obtain
 Two homologous groups of EEG $\boldsymbol{\widetilde{G}}_{i}^A = \{\widetilde{\boldsymbol{X}}^{s_k}_{v_i}|k=1,2,...,Q\}$ and $\boldsymbol{\widetilde{G}}_{i}^B = \{\widetilde{\boldsymbol{X}}^{s_k}_{v_i}|k=Q+1,Q+2,...,2Q\}$ can be obtained 
that sharing the similar group-level stimuli-related features. 
\par
Such data augmentation for group sample we represent it as follows function expression:

\begin{align}
\{\boldsymbol{\widetilde{G}}_{i}^A, \boldsymbol{\widetilde{G}}_{i}^B \}=Meiosis(\widetilde{\boldsymbol{G}}_{i})
\end{align}
\par
When Meiosis is built, for one minibatch of $P$ group samples $\mathcal{G}$, $2P$ group samples $\mathcal{\widetilde{G}}$ can be obtained as follows:
\begin{align}
\mathcal{\widetilde{G}} = \{\boldsymbol{\widetilde{G}}_{i}^t|i=1,2,...,P;t\in\{A,B\}\}=Meiosis(\mathcal{G})
\end{align}

$\boldsymbol{\widetilde{G}}_{i}^A$ could form a positive pair with $\boldsymbol{\widetilde{G}}_{i}^B$, form negative pairs with any other $2(P-1)$ group samples .

\subsection{Base Encoder}\label{AA}
To extract group-level stimuli-related features for contrastive learning, we fisrt design a base encoder to extract individual-level stimuli-related features from each individual EEG sample. We introduce the base encoder $f$ : $\mathbb{R^{M \times C}}$ $\rightarrow$ $\mathbb{R^{D}}$ which map individual EEG sample $\boldsymbol{X}$ to its representation $\boldsymbol{h}$ on a 512-dimensional feature space. Based on the existing model ResNet18-1D \cite{b20}, the base encoder is designed as follows:
\par
	As illustrated in Fig.4. It mainly contains 17 convolutional layers (Conv) with a 1D kernel. The kernels of the first convolutional layer parallel the time axis of the EEG signal tensor with a length of 9. Each residual block contains two convolutional layers with the same number and length of the kernels. In each residual block, kernels of the first layer parallel the time axis of the input EEG tensor, and the second layer parallels the channel axis. For the eight residual blocks, the length of the kernels is 15, 15, 11, 11, 7, 7, 3, and 3 in descending order. Max pooling with the 1D kernel (Maxpool), Avg pooling with the 1D kernel (Avgpool), Batch Normalization (BN), and Rectified Linear Unit (RELU) layers are shown in the corresponding positions in the figure.
\par
	Through the base encoder, for a augmented group sample $\boldsymbol{\widetilde{G}}_{i}^t$, its  individual-level stimuli-related representation set  \{${\boldsymbol{h}}_{1},{\boldsymbol{h}}_{2},...,{\boldsymbol{h}}_{Q}$\} can be obtained as by: 
	\begin{align}
\boldsymbol{H}_{i}^t = f(\boldsymbol{\widetilde{G}}_{i}^t)
	\end{align}
The set is used for further extracting group-level features. The individual representations can also be used for extracting emotional features for emotion classification.
\begin{figure}[htbp]
\centering
\includegraphics[scale=0.33]{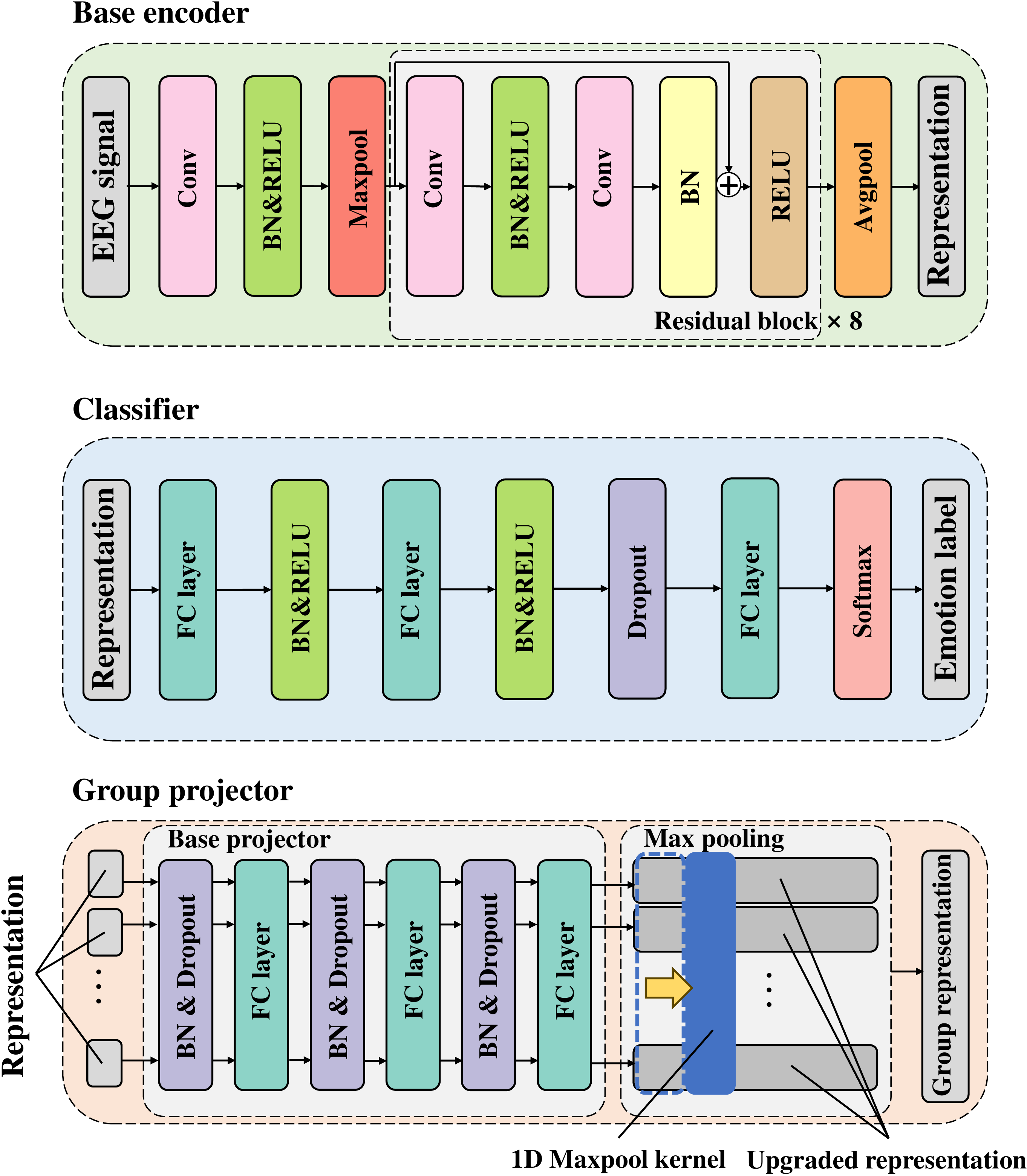}
\caption{Details of the architecture of the base encoder, group projector, and classifier. $Conv$ represent convolutional layer with 1D kernel. $Maxpool$ and $Avgpool$ represet Max pooling and Avg pooling with 1D kernel. $BN$ represent Batch Normalization. $FC$ layer represent fully-connected layer. $RELU$ represent Rectified Linear Unit.}
\label{fig_framework}
\end{figure}

\subsection{Group Projector}\label{AA}
	The group projector aims to accurately project stimuli-related representations into latent space from just 1-second EEG signals for calculating the similarity of video clip stimuli. To alleviate the hinders in extracting stimuli-related features from individual samples ( fatigue, distraction, etc), the group projector is designed to extract group-level features from multiple samples. 
\par
	A group of samples is an unordered set of matrixes that lacks a special extraction method. Most models focus on regular input representations. Such as the input of multi-channel images, there is a fixed order between different channels, as well as video, there is a fixed sequence between different frames. In the problem of unordered point cloud classification, \cite{b10} proposed PointNet adopting the symmetric function to build a network realized the features extraction of the unordered point cloud.

Inspired by it, we adopted a symmetric function to design a model suitable for extracting features from group EEG signals.
As illustrated in Fig.4 we designed the group projector consisting of a base projector and symmetric function MaxPool1D.
\par	
	To mitigate individual feature loss, the dimension of individual representation can be upgraded for extraction. We introduce the base projector $l$ : $\mathbb{R^{D }}$ $\rightarrow$ $\mathbb{R^{H}}$ that adopt a multilayer perceptron (MLP) to project each individual representation $\boldsymbol{h}$ on a 4096-dimensional feature space. The base projector contains three fully-connected layers with 1024, 2048, and 4096 hidden units in ascending order and adopt ReLU as  the activation function of the first two layers. Batch Normalization and Dropout with 0.5 are shown in the corresponding positions in the figure. 
\par

	To ensure an invariant output to represent the group sample with any input permutations,  1-dimension max-pooling (MaxPool1D) is adopted to aggregate the information from each dimension-upgraded representation. 
As illustrated in Fig.4, the 1D kernel of MaxPool1D is perpendicular to the dimension-upgraded representation vector. The scanning direction of the kernel is parallel to upgraded representation vector with a stride of 1, and the padding is 0. Such MaxPool can extract the maximum values on 4096 feature dimensions from $Q$ dimension-upgraded representations to obtain the group-level feature representation in latent space.
\par
 	We note group projector as $g$ : $\mathbb{R^{Q \times D}}$ $\rightarrow$ $\mathbb{R^{H}}$. 
Extracted group represetation in latent space can be obtained through $g$ as follows:
	\begin{align}
\boldsymbol{z}_v^t = g(\boldsymbol{H}_v^t)=MaxPool1D(l(\boldsymbol{h}_1),l(\boldsymbol{h}_2),...,l(\boldsymbol{h}_Q))
	\end{align}
\subsection{Classifier}\label{AA}
\par
	In the emotion classification fine-tuing task, we use the classifier to extract emotional features and predict emotion labels from the representations extracted by the base encoder. As illustrated in Fig.4 the classifier mainly contains three fully-connected layers with 512, 256, and 128 hidden units in descending order. Batch Normalization ReLU and Dropout with 0.5 are shown in the corresponding positions in the figure. 
\subsection{The Contrastive Loss}\label{AA}
\par

To measure the similarity of group-level stimuli-related features between two group samples, we can calculate the cosine similarity of their group representation vectors.
The input group samples $\{\boldsymbol{\widetilde{G}}_{i}^t|i=1,2,...,P;t \in\{A,B\}\}$ would be extracted to obtain group feature representations $\{\boldsymbol{z}_{i}^t|i=1,2,...,P;t \in\{A,B\}\}$ via the base encoder and group projector. Then, the similarity of two augmented group samples $\boldsymbol{\widetilde{G}}_{i}^A$ and $\boldsymbol{\widetilde{G}}_{j}^B$ can be calculated on $\boldsymbol{z}_{i}^A$ and $\boldsymbol{z}_{j}^B$:
	\begin{align}
	s(\boldsymbol{z}_{i}^A,\boldsymbol{z}_{j}^B)=\frac{\boldsymbol{z}_{i}^A\cdot\boldsymbol{z}_{j}^B}{\Vert \boldsymbol{z}_{i}^A\Vert \Vert \boldsymbol{z}_{j}^B\Vert},s(\boldsymbol{z}_{i}^A, \boldsymbol{z}_{j}^B)\in[0,1]
	\end{align}
\par
The contrastive loss is designed to maximize the similarity of two group-level representations of groups sharing the same stimuli label in a positive pair. Similar to the SimCLR framework\cite{b56}, we adopt the normalized temperature-scaled cross-entropy to define loss function as follows:
	\begin{small} 
	\begin{align}
	\ell_{i}^A\!=\!-\!log\frac{exp(\!s(\!\boldsymbol{z}_{i}^A, \boldsymbol{z}_{j}^B\!)\!/\!\tau\!)}{\sum_{\!j\!=\!1}^{\!P}\!\mathbb{1}_{\![j\!\neq\!i]\!}exp(\!s(\!\boldsymbol{z}_{i}^A,\boldsymbol{z}_{j}^A\!)\!/\!\tau\!)\!+\!\sum_{\!j\!=\!1}^{\!P}\!exp(\!s(\!\boldsymbol{z}_{i}^A,\boldsymbol{z}_{j}^B\!)\!/\!\tau\!)}
	\end{align}
	\end{small} 
where $\mathbb{1}_{[j \neq i]} \in \{0,1\}$ is an indicator function equaling to 1 if  $j \neq i$. 
$\tau$ is the temperature parameter of softmax. The smaller the loss function is, the larger similarity between $\boldsymbol{z}_{i}^A$ and $\boldsymbol{z}_{i}^B$ , and the smaller the similarity between $\boldsymbol{z}_{i}^A$ and other group representations come from the same minibatch. % pairs involving $\boldsymbol{z}_{i}^A$.
\par
Finally, the total loss for an iteration is the average of all contrastive losses for backpropagation as follows:
	\begin{align}
\mathcal{L}=\frac{1}{2P}\sum_{i=1}^P(\ell_{i}^A+\ell_{i}^B)
	\end{align}

\subsection{Pre-training Process}\label{AA}
	Based on the constructed group sampler, data augmentation, base encoder, group projector, and loss function the SGMC pre-training can be performed.
\par
	%Training would be performed in thousands of epochs until the model learned the best representation.
In a pre-training, we first set a number of epochs $T_1$, and then iterate the epoch.
In each epoch, we continue to sample $P$ video clips per iteration until all video clips are enumerated. 
Each iteration, Sampler extract $2PQ$ EEG samples $\mathcal{D}=\{\boldsymbol{X}^{s_{k}}_{v_{i}}|i=1,2,...,P;k=1,2,...,2Q\}$ and pack them into groups $\mathcal{G}=\{\boldsymbol{G}_{i}|i=1,2,...,P\}$.

Nextly for the Meiosis data augmentation, to avoid the model cheating by recognizing the split position, we randomly generate a fixed split position $c$, sent it to each time of Meiosis in this iteration ($1<c<M-1$). $2Q$ augmented group samples $\widetilde{\mathcal{G}}=\{\widetilde{\boldsymbol{G}}_{i}^t|i=1,2,...,P;t\in\{A,B\}\}\}$ can be obtained by (3). Further we extract group-level features and project them to latent space to obtain group representations by (4) and (5).
Furthermore, we calculate loss  $\mathcal{L}$ by (6)-(8). Finally, we abate loss $\mathcal{L}$ by backpropagation to calculate the gradient for optimizer updating parameters of $f$ and $g$. Detailed procedures are summarized in Algorithm 1.

\begin{algorithm}[H]
    \caption{Self-supervised Group Meiosis Contrastive Learning}
    \begin{algorithmic}[1] 
        \REQUIRE Number of video clips $P$ per minibatch, number of subjects $Q$ per group. Initilized base encoder $f$ and group projector $g$.

        \FOR {{$epoch=1$ to $T_1$}}
				\REPEAT
				\STATE Sample $P$ video clips $\{v_i|i=1,2,...,P\}$. 
				\STATE Randomly select $2Q$ subjects $\{s_k|k=1,2,...,2Q\}$. 
				\STATE Sampler pack minibatch $\mathcal{G}=\{\boldsymbol{G}_{i}|i=1,2,...,P\}\}$ from $\mathcal{D}=\{\boldsymbol{X}^{s_{k}}_{v_{i}}|i=1,2,...,P;k=1,2,...,2Q\}$
				\STATE Randomly generate a split position $c$. 
				\STATE Obtain $\widetilde{\mathcal{G}}=\{\widetilde{\boldsymbol{G}}_{i}^t|i=1,2,...,P;t\in\{A,B\}\}$ from $\mathcal{G}$ through Meiosis with $c$ by (1)-(3). 
				\STATE Obtain $\mathcal{Z}=\{\boldsymbol{z}_{i}^t|i=1,2,...,P;t=\in\{A,B\}\}$ from $\widetilde{\mathcal{G}}$ through $f$ and $g$ by (4) and (5). 
				\STATE Calculate loss $\mathcal{L}$ by (6)-(8). 
				\STATE Abate loss $\mathcal{L}$ through optimizer updating parameters of $f$ and $g$.  
				\UNTIL{all video clips are enumerated.}
        \ENDFOR
        \ENSURE base encoder $f$, throw away group projector $g$.
    \end{algorithmic} 
\end{algorithm}
%%%%%%%%%%%%%%%%%%%%%%%%%%%%%%%%%%%%%%%%%
\subsection{Fine-tuning Process}\label{AA}
\par
	To achieve excellent emotional classification performance, based on learned feature representations we further fine-tune the model with labeled samples. As illustrated in Fig.1 emotion classification supervised training is performed on the model consisting of an initialized classifier and the SGMC pre-trained base encoder. 
\par
	We denote the training data as $\boldsymbol{X}$ and their labels as $\boldsymbol{y}$. We denote the classifier as $k(\cdot)$ The label $\boldsymbol{y}$ is a categorical variable. For example, if there are four emotional categories, $\boldsymbol{y}$ can take four values: 0, 1, 2 or 3. We need to predict the emotion category $\boldsymbol{y}$ for each sample $\boldsymbol{X} \in \mathbb{R^{M \times C}}$. The pre-trained base encoder $f$ extracts the representation from original EEG signal $\boldsymbol{X}$ for classifier $k(\cdot)$ extract predictive features to obtain prediction category $\boldsymbol{y^{pre}}=k(f(\boldsymbol{X}))$. We apply the cross entropy function to define the loss function for the emotion classification task and apply an optimizer to minimize the loss function to optimize the parameters of the model. 
%Based on the SGMC pre-trained base encoder that has learned feature representation.
Finally, when the loss function converges, a predictive EEG-based emotion recognition model is obtained.

\section{Experiments}
In this section, we introduce the implementation detail on the DEAP and SEED dataset and our experiment evaluation. In our experiment, we verify the effectiveness by comparing the SGMC with other competitive methods of emotion recognition and evaluating its performance on limited labeled sample learning. Further, we explore the reason for the effectiveness by visualizing the feature representation learned by the SGMC. Moreover, we explore the meaningful law of the framework by evaluating the different combinations of hyper parameters. %and evaluating the different combinations of hyper parameters.
Furthermore, we verify the rationality of architecture design by conducting control and ablation experiments. 
\\
%\textbf{Emotion Classification Performance}
\subsection{Implementation Detail}
In this section, we elaborate on our implementation detail of the dataset, data processing, and basic hyper parameters utilized in the experiments.
\\
\noindent
%\subsection{Emotional EEG Databases}\label{AA}
\begin{table*}[b]
\fontsize{9}{10}\selectfont
\centering
\caption{ Hyper parameters utilized in the proposed SGMC }
% 在tabular 上面引入threeparttable环境
\begin{threeparttable}
 \begin{tabular}{ll|lllllllll}  
\toprule   
 % & & (1)&(2)&(3)&(4)&(5)&(6) \\ 
   \multicolumn{2}{c|}{}    & $Epoch$ &$batchsize$&$lr$&$\tau$&$P$&$Q$&$Shape_{tr}$&$ Shape_{te}/Shape_{val}$ \\ 
%\cline{1-11}
\toprule
% & &Y&Y&Y&Y&N&\\
%\midrule   
 %Baseline & &  &\textbf{100.00} &00.00&00.00&00.00\\
\multirow{2}{*}{DEAP}&Pre-training&2800 & 32&$10^{-4}$&$10^{-1}$& 8 & 2& $(1680,32,1,32,128)$&$(360,32,1,32,128)$  \\
%\cline{2-2}
\cmidrule(lr){2-2} 
    &Fine-tuning &60& 2048&$10^{-3}$& - &-&-&$(53760,1,32,128)$& $(11520,1,32,128)$ \\
%\cline{1-11}
\toprule 
\multirow{2}{*}{SEED}&Pre-training&3288 & 64&$10^{-3}$&$10^{-1}$& 16 & 2& $(2374,45,1,62,200)$&$(510,45,1,62,200)$  \\
%\cline{2-2}
\cmidrule(lr){2-2} 
    &Fine-tuning &70& 256&$10^{-3}$& - &-&-&$(106380,1,62,200)$&$(22950,1,62,200)$  \\
\midrule   
 %SeqCLR[]&&& 00.00& 00.00&00.00&00.00\\
%  \bottomrule  
\end{tabular}
% 这个地方是引入注释或者标注
\begin{tablenotes}
\item $ Shape_{tr},Shape_{te}, Shape_{val}$ respectively represent size of tensor of training test and validation dataset for pre-training or fine-tuning. $Epoch$ represent an appropriate number of the pre-training or fine-tuning epochs for achieving the fine emotion recognition performance. $batchsize$ represent the number of samples in a minibatch. %$Q$ represent the number of samples per group. $P$ represent the number of selected video clip in a minibatch.
%\item[2] B represent C
\end{tablenotes}
\end{threeparttable}
\label{tabXXXxXXXXXXXX}
\end{table*} 
(1) Dataset
	\par
DEAP:
%	The widely-used DEAP dataset recorded 32-channel EEG signals and 8-channel peripheral physiological signals of 32 subjects when watching 40 one-minute-long music videos. 
	The widely-used DEAP dataset \cite{b41} includes 32-channel EEG signals and 8-channel peripheral physiological signals recorded by 32 subjects when watched 40 pieces of a one-minute music video. Each trial data was recorded under 3-seconds of resting state and 60-seconds of stimuli. The recorded EEG signals are down-sampled to a 128 Hz sampling rate and processed with a bandpass frequency filter from 4-45 Hz by the provider. % Here the EEG signals were sampled at 512 Hz and then down-sampled to 128 Hz.
%After watching each video, participants rate their arousal levels, valence, liking, and dominance from one to nine for each video. 
After watching each video, subjects were asked to rate their emotional levels of arousal, valence, liking, and dominance from 1 to 9 for each video. We adopt the EEG signals and rating values of arousal and valence to perform emotion recognition.
%The participants were asked to rate their levels of arousal, valence, liking, and dominance for each video from 1 to 9.
%The EEG signals and rating values are utilized to construct the emotion recognition task in our experiments. 
%In terms of the emotional rating value of each trial in the range of $1.0$ to $9.0$ in the arousal and valence domains, the median of 5.0 was used as the threshold to divide the rating value into two categories.
We set the threshold value of the rating value of arousal and valence at 5. When the rating value is more than 5.0, the corresponding EEG signals are labeled as high arousal or valence. Otherwise, it is labeled as low arousal or valence.
% In the cases where the emotional rating value is rated more than 5.0, the corresponding EEG signals are labeled as high arousal or valence. On the contrary, for the ones less than or equal to 5.0. 
%Given EEG signals to predict corresponding labeled categories, the emotion classification tasks can be formulated as arousal binary classification valence binary classification, and arousal valence four classification problems in this paper.
Each EEG signal corresponds to valence and arousal two labels, which can be used to construct two or four classification tasks.
%\par
	%In our experiment, 
	
\par
SEED:
	The SEED dataset is widely used in emotion recognition algorithms \cite{b42}. 
The dataset recorded the EEG signals from 15 subjects when watching 15 videos selected from movies in three categories of emotions, including positive, neutral, and negative. Each video is about 4 minutes long.
%The dataset contains the EEG data recorded from 15 subjects. Each of them watched 15 videos chosen from movies in three categories of emotions, including positive, neutral, and negative.
% Each video was selected to elicit a single desired target emotion. The length of each video is about four minutes. 
Each subject repeated the experiments for three sessions, with an interval of more than one week.
%Each subject was required to carry out the experiments for three sessions. There was a one-week or longer time interval between two sessions. 
%For each session, the subjects watched one video for each trial, resulting in 15 trials. 
%EEG signals were recorded using an ESI NeuroScan System2 with 62 electrodes placed according to the international 10-20 system with a sampling rate of 1000 Hz and have been downsampled to 200 Hz and filtered from 0 to 75 Hz.
The EEG signals were recorded via 62 electrodes at a sampling rate of 1000Hz and have been downsampled to 200 Hz and filtered from 0 to 75 Hz by the provider.\\
(2) Data Process 
\par
On the DEAP, we use a 1-second-long sliding window to separate the 63s signal of each trial into 63 non-overlapping EEG signal segments. %Futher we use a non-overlapping sliding window to separate the trial data into one-second-long signals. For the next step, 
To improve accuracy, following existing work \cite{b28} we reduce the 3s resting state EEG signals from the 60s emotional stimuli EEG signal. In detail, in each trial, we average the 3s baseline EEG signal segments to get a 1s average baseline EEG signal segment. The remaining 60 segments each subtract the average baseline segment to become input samples. %Finally, we obtain 76800 windows from data of 32 subjects watching 40 60-second-long videos. 
All samples correspond to a total of 2400 (40 videos with 60-seconds-long) repeated 1-second-long video clips. 
%The training set, testing set, and validation are randomly divided according to the video clips which are separated by sliding window in the ratio of 70:15:15. 
%We randomly divide 1680 clips and their corresponding 53760 samples into training sets, 320 clips and their corresponding 11520 samples into test sets, and 320 clips and their corresponding 11520 samples into verification sets according to the ratio of 70:15:15. 
1680, 320, and 320 1-second video clips are randomly divided into three sets from 2400 video clips in the ratio of 70:15:15. %for the training set, testing set, and validation set respectively. 
These three sets of video clips that were watched by 32 subjects correspond to 53760, 11520, and 11520 (70:15:15) EEG segments which are used as the training set, testing set, and validation set respectively.
%53760, 11520, 11520 EEG segments recorded by 32 subjects who watching that corresponding to such 1680, 320, and 320 video clips are divided into the training set, testing set, and validation set respectively.
%The 53760, 11520, and 11520 EEG segments  are divided into the training set, testing set, validation set
%The 53760 EEG segments recorded by 32 subjects watching 1680 
\par
On SEED, we first perform an L2 normalization for each trial of EEG signal in each channel. Similar to the DEAP dataset we divide movie videos into 1-second windows. Because the length between the trial videos is different, we segment adjacent windows from front to back according to the time axis until the coverage of windows exceeds the video range. 3394 video clips are obtained from 15 movie videos and randomly divided into 2734, 510, and 510 clips   
, which three sets of video clips are in the ratio of 70:15:15.
%for the training set, test set, and validation set in the same ratio of 70: 15: 15.  The 123030, 22950, 22950 EEG segments recorded by 15 subjects in three sessions corresponding watching such 2734, 510, 510 video clips are divided into the training set, testing set, and validation set.
These three sets of video clips that were watched by 15 subjects three times correspond to 123030, 22950, and 22950 (70:15:15) EEG segments which are used as the training set, testing set, and validation set respectively.
\\
(3) Basic Configuration
\par
To accurately evaluate the performance of emotion recognition for a pre-training framework, there are two steps we adopted for evaluating the results. 
We first save pre-trained models with the different epochs. Next, we select the model with the highest average accuracy on emotion recognition obtained by five times of fine-tuning. 
 Such average accuracy is evaluated as the result.% Such the corresponding number of epochs for pre-training is the most suitable and is represented as $Epoch_{pre}$.
\par
To speed up sampling, in the pre-training process we set the five axes of dataset tensor to correspond to $video$ $clip$, $subject$, $1$, $channel$, $sampling$ $point$ respectivey. In the fine-tuning process, the first two axes $video$ $clip$ and $subject$ of the dataset are reshaped into a sample axis. Each axis of reshaped dataset corresponds to $sample$, $1$, $channel$, $sampling$ $point$ in turn. In the pre-training task, each epoch traverses every video clip of the dataset, a fine pre-training task generally needs to train more than 2000 epochs. To reduce the workload, we use the validation dataset to adjust the hyper parameters of the SGMC framework and use the test dataset to evaluate the model.
The tensor shape of the training set, testing set, and validation set are represented as $Shape_{tr}$, $ Shape_{te}$, and $ Shape_{val}$ and are listed in Table \uppercase\expandafter{\romannumeral1}. 
%We represent $Shape_{tr}$ as the size of the tensor of the training set, represent $ Shape_{te}, Shape_{val}$ as the size of the tensor of the test and validation set. There are five axes of the dataset in pre-training tasks corresponding to (video clip, subject, 1, channel, sampling point).
\par
	In this paper, we use PyTorch \cite{b29} to implement our experiments based on the NVIDIA RTX3060 GPU. 
The Adam optimizer \cite{b30} is used to minimize the loss functions for both the pre-training and fine-tuning process. We represent $lr$ as the learning rate of the optimizer.
 In the pre-training process and fine-tuning process, the number of epochs, batch size, the temperature parameter $\tau$, learning rate $lr$, number of video clips per iteration $P$,  number of samples per group $Q$, and size of the tensor of the dataset have applied different values, as shown in  Table \uppercase\expandafter{\romannumeral1}, we list all hyper parameters utilized in two processes on DEAP and SEED dataset.
\subsection{Emotion Classification Performance}
\noindent
(1) Performance on DEAP
\par
%\begin{table}[H]
%\fontsize{9}{10}\selectfont
%\centering
%\caption{ SEED dataset}
% SEED超参
%\setlength{\tabcolsep}{8.5mm}{
%\begin{threeparttable}
% \begin{tabular}{ll}  
%\toprule   
%  Method & Accuracy($\%$)  \\ 
%\midrule   
 %CNN-LSTM (2020)[] & 90.82 & 86.13 & - \\
 %CDCN (2020)[] & 92.24 & 92.92 & - \\
 %MMResLSTM (2019)[]&92.87& 92.30& -\\
 %MCLFS-GAN (2020)[] & - & - & 81.32\\
 %GANSER (2021)[] & 93.52 & 94.21 & 89.74 \\
 %BiDANN(2018)&92.38\\
 %BiHDM(2019) &93.12\\
 %Resnet18 1D kernel(2021)&93.42\\
 %SeqCLR
 %Baseline(Fully-supervised) & 87.68 \\
 %Proposed(GSCL) & \textbf{94.04}   \\
% Ours($q=4$) & 48.27 & 0.34 & -\\
  %\bottomrule  
%\end{tabular}
% SEED比较结果表
%\begin{tablenotes}
%\item  Average ACC($\%$) of state-of-the-art method on \textbf{SEED} dataset for positive neutral and negative three classification .
%\item
%\end{tablenotes}
%\end{threeparttable}
%\label{tabXXX}}
%\end{table}
\par
	As illustrated in Table \uppercase\expandafter{\romannumeral2}, 
On the DEAP dataset,  We first compare the SGMC with four state-of-the-art methods in the two emotion dimensions of valence and arousal: one residual long short-term memory network utilizing multi-modal data MMResLSTM \cite{b32}, a channel-fused dense convolutional network CDCN \cite{b31}, and a hybrid network of convolutional neural networks and recurrent networks with a channel-wise attention mechanism  ACRNN \cite{b43}. From Table \uppercase\expandafter{\romannumeral2}, it can be found that the accuracy of the proposed SGMC is 1\% higher than the second in the valence dimension and 2.3\% higher than in the arousal dimension. The comparison results demonstrate the effectiveness of the SGMC on EEG-based emotion recognition.
\par
%    To verify the effectiveness of our method in the self-supervised learning field, we further compare the proposed method with a GAN-based data augmentation SSL method MCLFS-GAN \cite{b34}  GANSER \cite{b35}. 
To verify the effectiveness of the proposed framework in the data augmentation and self-supervised learning fields, we further compare the SGMC with a GAN-based data augmentation method MCLFS-GAN \cite{b34} and a self-supervised GAN-based data augmentation framework GANSER \cite{b35}.
Especially, according to the experimental setting of  MCLFS-GAN \cite{b34} and GANSER \cite{b35}, we further performe a comparison on a four-category classification problem: distinguishing EEG signals of four categories: high valence and high arousal, high valence and low arousal, low valence and high arousal, and low valence and high arousal. In Table \uppercase\expandafter{\romannumeral2}, it can be found that the proposed method outperforms the existing data augmentation and self-supervised learning method over $11.33\%$ and $2.09\%$ on four-category classification. As illustrated in Fig.5. meanwhile, the confusion matrices of the SGMC on four-category classification are presented. It shows that the SGMC achieves good performance in each category, especially in low arousal and high valence. % We can also find that even in the formulation of four-category classification, our proposed method can correctly classify nearly $94\%$ EEG signals at valence and arousal dimensions. 
\par 
Furthermore, we first compare the proposed SGMC with our own fully-supervised baseline using the same network model without pre-training. In valence, arousal, and four-category dimensions, the accuracy of the SGMC exceedes the fully-supervised baseline over $3.49\%$ $3.32\%$ and $4.97\%$, which shows the significant effect of improving emotion recognition. 
%As illustrated in Fig.5. The confusion matrices of the SGMC on four-category classification is presented, which shows our method can better recognize low arousal and high valence, low arousal and low valence than high arousal and low valence, high arousal and high valence. Each categorie achieve a good accuracy over $91.7\%$.\\
\begin{table}%[H]
\fontsize{9}{10}\selectfont
\centering
\caption{ Performances on DEAP }
% DEAP超参
\setlength{\tabcolsep}{3.2mm}{
\begin{threeparttable}
 \begin{tabular}{llll}  
\toprule   
  Method & Valence &  Arousal & Four\\ 
\midrule   

 CNN-LSTM (2020)\cite{b28} & 90.82 & 86.13 & - \\
 CDCN (2020)\cite{b31} & 92.24 & 92.92 & - \\
 MMResLSTM (2019)\cite{b32}&92.87& 92.30& -\\
 ARCNN (2019)\cite{b43}&93.72& 93.38& -\\
\midrule   
 MCLFS-GAN (2020)\cite{b34} & - & - & 81.32\\
 GANSER (2022)\cite{b35} & 93.52 & 94.21 & 89.74 \\
\midrule   
 Proposed(Fully-supervised) & 91.23 & 92.36 & 87.68\\
 Proposed(Fine-tuned) & \textbf{94.72} & \textbf{95.68} & \textbf{92.65} \\
% Ours($q=4$) & 48.27 & 0.34 & -\\
  \bottomrule  
\end{tabular}
% DEAP比较结果表
\begin{tablenotes}
\item Average accuracy($\%$) of state-of-the-art method on the DEAP dataset for valence classification, arousal classification and four classification.
\item
\end{tablenotes}
\end{threeparttable}
\label{tabXX}}
\end{table}
\begin{figure}%[H]
\centering
\includegraphics[width=0.45\textwidth,height=0.37\textwidth]{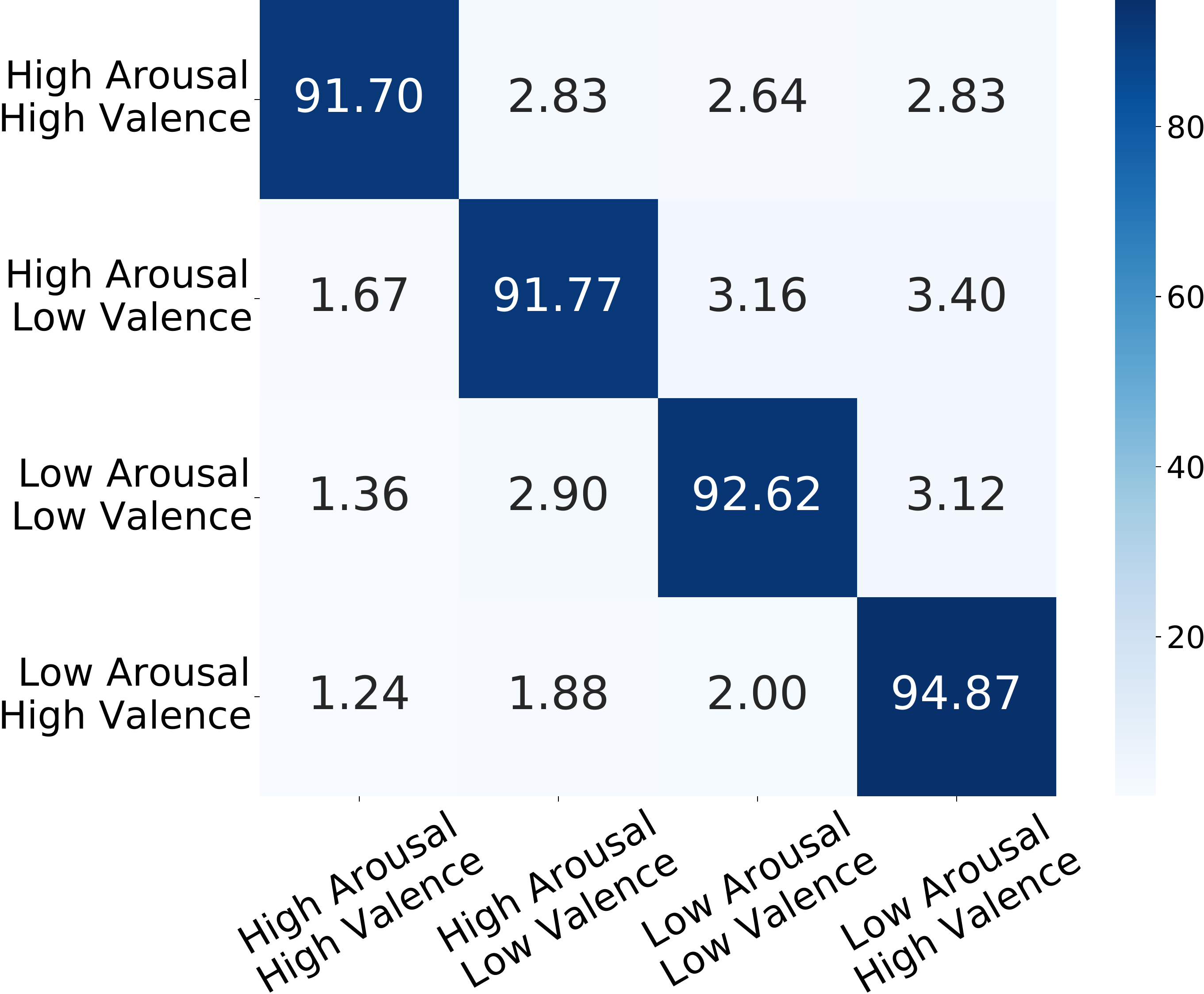}
    \caption{The confusion matrix of classification on DEAP}
\end{figure}
%As illustrated in Fig.5. The confusion matrices of the SGMC on four-category classification is presented. It shows that the SGMC achieves good performance in each category, especially in low arousal and high valence.
\noindent
(2) Performance on SEED

\par
	As illustrated in Table \uppercase\expandafter{\romannumeral3},
Similar to the DEAP,
  we first compare our proposed SGMC with four fully-supervised state-of-the-art studies %on the full labeled training set
: GRSLR \cite{b51} adopting a graph regularized sparse linear regression model, BiHDM \cite{b39} utilizing two independent recurrent networks for the left and right hemispheres of the brain, DGCNN \cite{b38} adopting a dynamic graph convolutional neural network, and a 1D CNN-based residual neural network ResNet18 \cite{b20}. Results use accuracy in the three classification tasks of positive neutral, and negative emotions. As illustrated in Fig.5. The details of the classification result are shown in the confusion matrix. The SGMC achieves good accuracy in three categories, especially performing better on positive than negative and neutral.
%obtains higher accuracy on positive than neutral and negative, especially neutral and negative are more easily confused. 
As illustrated in Table \uppercase\expandafter{\romannumeral3} the proposed SGMC outperforms the four state-of-the-art studies, reflecting its good emotion recognition performance on the SEED.%on a larger dataset.
\par
%%%%%%%%%%%%%%%%%%和自监督基线比先去掉

%%%%%%%%%%%%%%%%%%%
\par
Further, we compare the SGMC with our fully-supervised baseline using the same model. Especially, the SEED dataset has nearly five times the data volume of the DEAP dataset. Therefore, it can better reflect the performance of self-supervised learning by utilizing a large number of unlabeled samples to make up for scarce artificial accurate labels. We report results obtained from fine-tuning with four various percentages of the total training set labeled samples (based on pre-training on the full training set). From $1\%$ to $50\%$ percentage of labeled samples, the SGMC exceeds our fully-supervised baseline over $44.84\%$, $33.52\%$, and $8.24\%$. Such results show the proposed SGMC can take advantage of consistency of stimuli to significantly make up for artificial accurate labels. Using the full training set labeled samples to fine-tune, the SGMC significantly exceed our fully-supervised baseline over $4.27\%$ as well. This shows the SGMC contributes a significant improvement by utilizing large unlabeled data.
\\

\begin{table}%[b]
\fontsize{9}{10}\selectfont
\centering
\caption{ Performances on SEED }%Result of various symmetry function for group projector on downstream performance on \textbf{SEED} dataset. Result is average of five times of emotion classification training }
% 在tabular 上面引入threeparttable环境
\setlength{\tabcolsep}{0.55mm}{
\begin{threeparttable}
 \begin{tabular}{l|llllll}  
\toprule   
 % & & (1)&(2)&(3)&(4)&(5)&(6) \\ 
   \multicolumn{1}{c|}{Method}&\multicolumn{4}{c}{Accuracy($\%$)}\\
\midrule   
%\cline{1-5}
   \multicolumn{1}{c|}{Percentage of labels}    & $1\%$ &$10\%$&$50\%$&100\%& \\ 
% & &Y&Y&Y&Y&N&\\
\midrule   
 %Baseline & &  &\textbf{100.00} &00.00&00.00&00.00\\
 GRSLR(2018)\cite{b51} &-&-&-&87.39\\
 DGCNN(2018)\cite{b38}&-&-&-&90.40\\
 BiHDM(2019)\cite{b39}&-&-&-&93.12\\
 ResNet18 1D kernel(2021)\cite{b20}&-&-&-&93.43\\
\midrule    
 Proposed(Fully-supervised)&44.81& 59.77&85.47& 89.83   \\
 %Proposed(Frozen)   &\textbf{89.65}& \textbf{93.29}&\textbf{93.71}&\textbf{94.04}     \\
 Proposed(Fine-tuned)   &\textbf{89.65}& \textbf{93.29}&\textbf{93.71}&\textbf{94.04}     \\
 %MaxPool1d   &\textbf{46.12}& \textbf{75.46}&\textbf{89.57}&\textbf{92.65}     \\
\midrule   
 %SeqCLR[]&&& 00.00& 00.00&00.00&00.00\\
%  \bottomrule  
\end{tabular}
% 这个地方是引入注释或者标注
\begin{tablenotes}
\item Average accuracy($\%$) of state-of-the-art method on the SEED dataset for  postive, neutral and negative three-classification. Percentages of labels represent labeled samples use to training emotion recognition account for the percentage of the full training set.% * Represent the experiment results are obtained based on our own implementation.
\item
\end{tablenotes}
%\begin{tablenotes}
%\item%[1] A represent B
%\item%[2] B represent C
%\end{tablenotes}
\end{threeparttable}
\label{tabXXXxXXXXXXXX}}
\end{table}
\begin{figure}%[b]
\centering
\includegraphics[width=0.42\textwidth,height=0.35\textwidth]{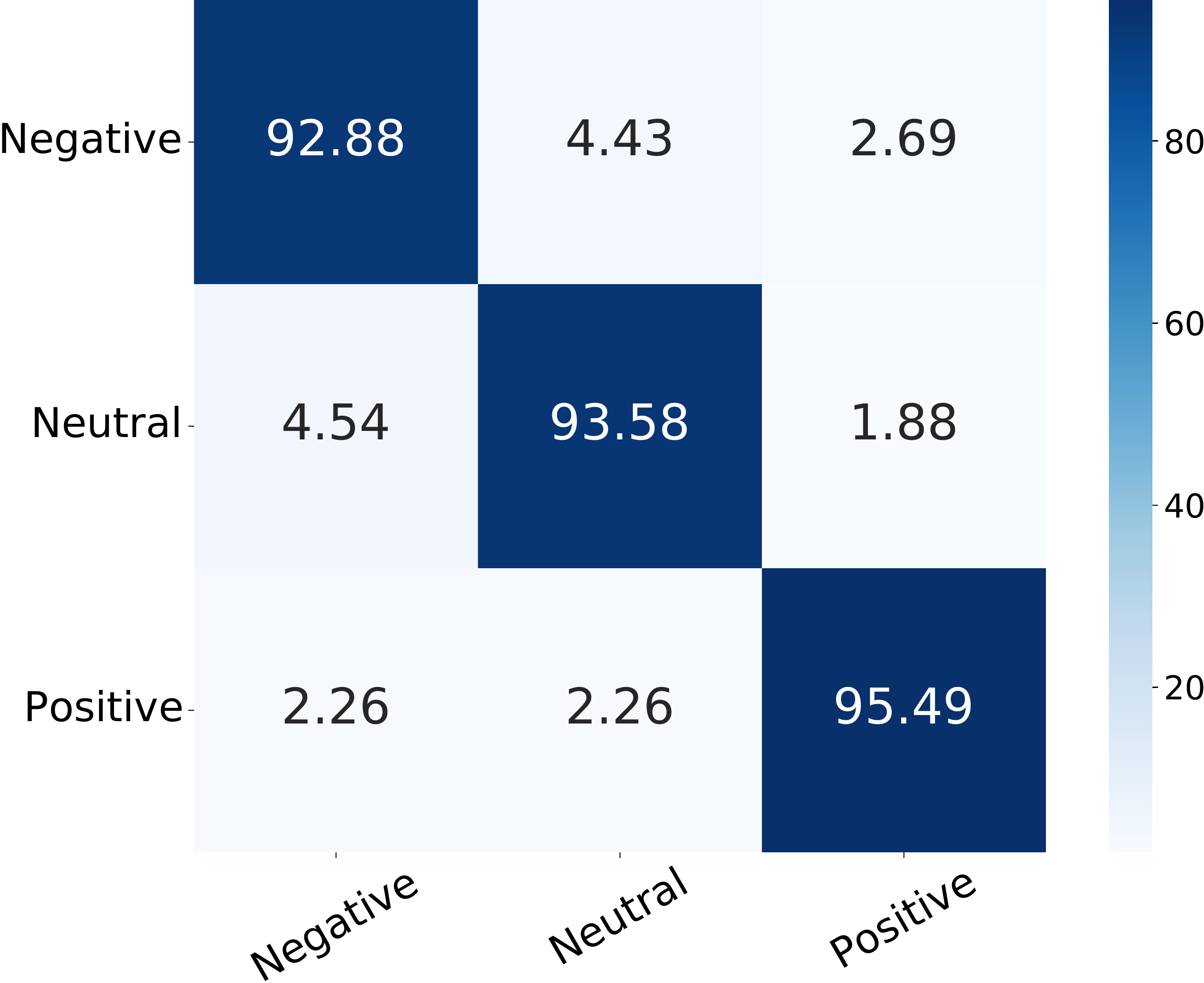}
    \caption{The confusion matrix of classification on SEED}
\end{figure}
\begin{figure*}%[b]
\centering
\includegraphics[width=0.95\textwidth,height=0.5\textwidth]{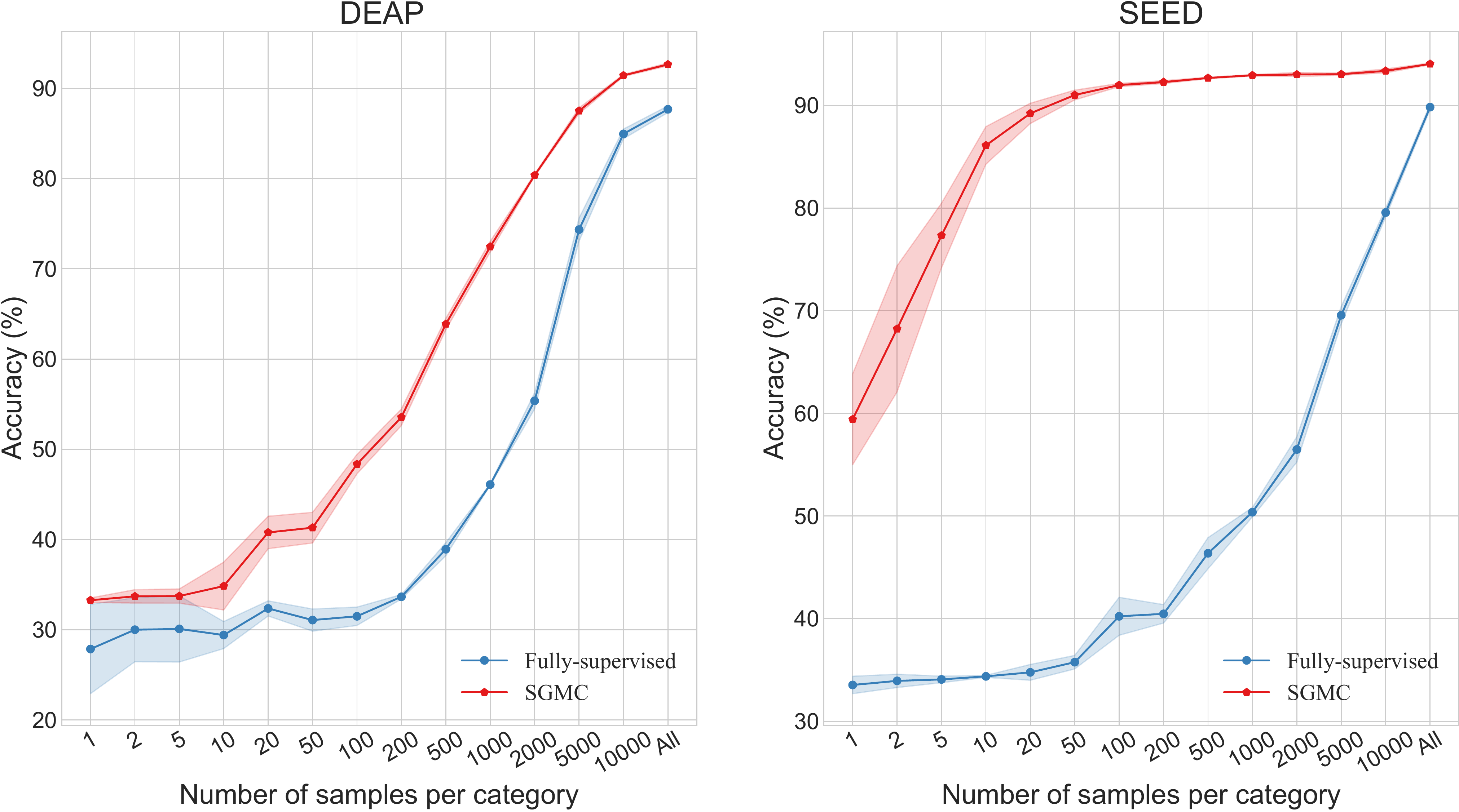}
\caption{ 
Learning effect of the labeled sample size for the emotion recognition. the left is for the result on DEAP, and the right is for SEED. %Red lines are the average accuracy with five times of the SGMC fine-tuning, and the blue lines are the average accuracy with fully-supervised training for five times.
The red line represents the model is based on the SGMC pre-trained with the full training set and fine-tuned with the different number of labeled samples per category. Blue line represents the model is only fully-supervised trained with the different number of labeled samples per category. The results are the average test accuracy of five times of emotion classification training and the shade area represents standard deviation.}
\label{fig_framework}
\end{figure*}
\noindent
\subsection{Performance on Limited Labeled Sample Learning}
%2)Performance on limited labeled data
\par

Based on the above results on SEED, it can be found that fewer labeled samples can also lead to good results. To evaluate the performance on limited labeled sample learning, we further evaluate the results on DEAP and SEED when the number of labeled samples per category increasing. %, and observe the changing trend of the results.
 We adopte a model based on SGMC pre-trained with the full training set and an initialized model to compare their performance on fine-tuning/fully-supervised learning with the same limited labeled sample.
On the DEAP the results adopt a four-category classification of arousal and valence. On SEED the results adopt a three-category classification.
As illustrated in Fig. 7. the results of the different number of labeled samples per category for fine-tuning/fully-supervised learning are reported.
\par
we can find that in any amount of labeled samples regimes, the accuracy of the SGMC fine-tuning is significantly superior to the fully-supervised baseline, and it is more significant in the lower labeled samples regime. 
On the DEAP dataset, when the number of labeled samples per category is over 10, the performance of the SGMC significantly outperforms the supervised. When fine-tuned with 5000 labeled samples per category (37.2\% of the full training set), the SGMC reaches a good accuracy of $87.51\%$ which is nearly by $87.68\%$ of fully-supervised accuracy training with the full training set.
On the SEED dataset,
when fine-tuned with even only one labeled sample per category ($0.00278\%$ of the training set), the SGMC reached an accuracy of $59.42\%$. 
When fine-tuned with 50 samples per category ($0.14\%$ of the training set), the accuracy of 91.01\% outperforms the fully-supervised baseline with 100\% labeled data. Further, we observe that when the number of category is over 500, the curve has converged. This shows the SGMC enables a significant decline in the demand for artificial labels and reflects the consistency of stimuli have been well exploited to make up for artificial labels.
%This reflects that the potential stimuli labels have been well exploited to make up for artificial labels.\\
\subsection{Representation Visualization}
\par
To explore how SGMC contributes to superior performance on emotion recognition, we visualize the learned feature representations of the SGMC fine-tuned model and the only fully-supervised model.
\begin{figure}%[H]
\centering
\includegraphics[width=0.45\textwidth,height=0.42\textwidth]{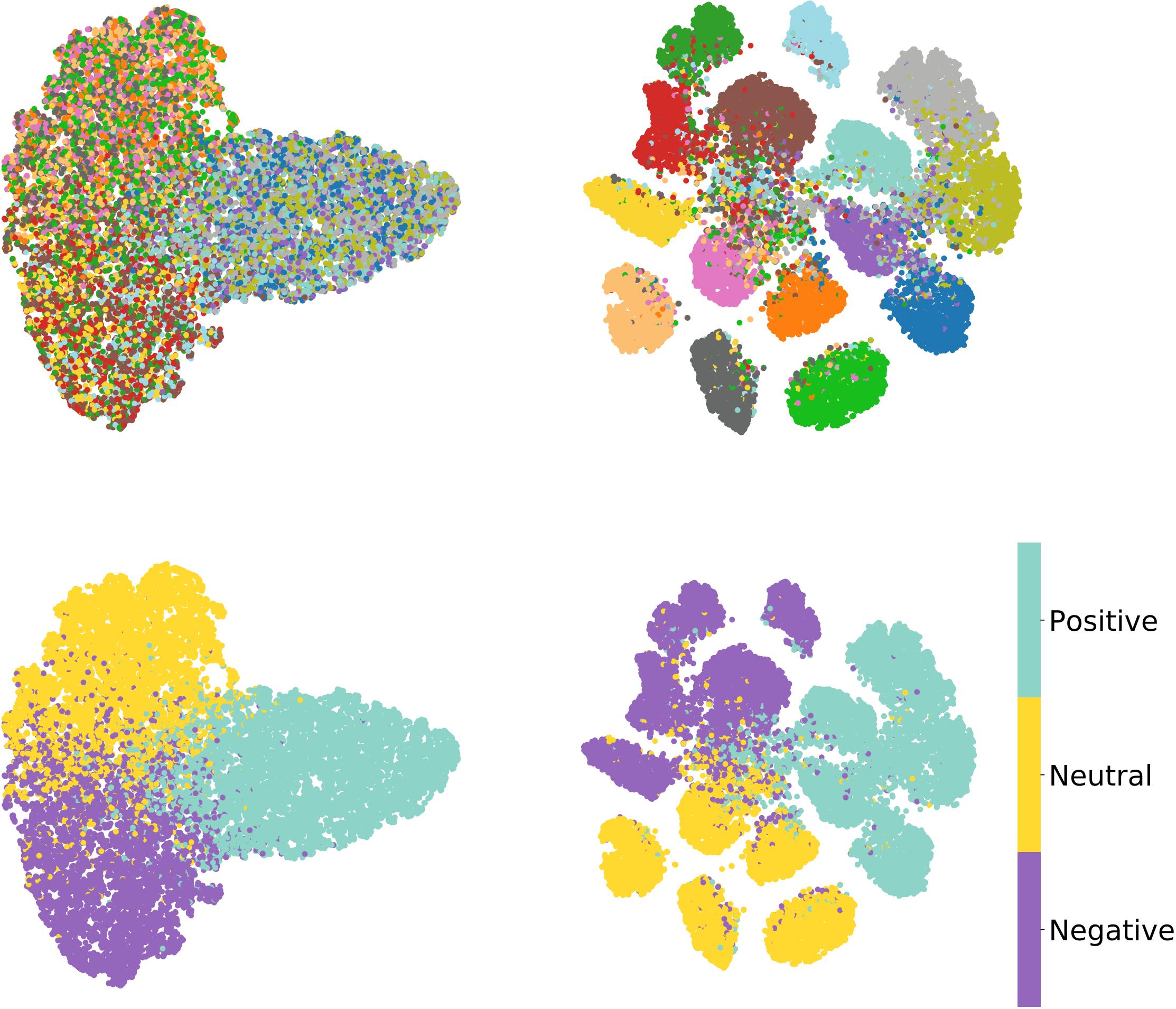}
\caption{t-SNE visualization for feature representations demonstrated on SEED with fully-supervised (left) and the SGMC fine-tuned (right). Tops are the visualization marked by movie videos, and the different colors represent the 15 videos. Bottoms are the visualization marked by emotion labels, and three colors represent positive, neutral, and negative videos respectively.
} \label{fig1}
\end{figure} 
\par
As illustrated in Fig. 8, the 512-dimension feature representations extracted by the base encoder from the samples of the full SEED testing set are projected to two dimensions through t-SNE\cite{b52}. In the figure above, 15 colors represent samples corresponding to the 15 trial video clips (about 4-minutes). It can be found that in the visualization of the SGMC fine-tuned (right), the feature representation of the same video clip tends to gather together to form 15 distinguishable groups.  On the contrary, in the visualization of fully-supervised (left), the representations corresponding to the different video clips cannot be distinguished significantly. Visualization reveals that the SGMC not only learns stimuli-related feature representations but also enables the model to distinguish whether different stimuli come from a continuous video.
%This shows that the SGMC enables the model to learn video-level representation.\par
	Further, we mark the corresponding emotion labels with three colors in the figure below. There are more indistinguishable representations with different emotion labels mixed together in fully-supervised visualization     (left). In the SGMC fine-tuned visualization (right), there are fewer feature representations with the different emotion labels mixed together and shows better emotional discrimination. It reflects that the SGMC enables the model to learn the video-level stimuli-related representation to improve emotion recognition performance.

\par

\subsection{Effect of Hyper Parameters}
\par
To explore the effect of the number of samples per group (Q) and the number of selected video clips per iteration (P) on contrastive learning, we evaluate various combinations of hyper parameters. 
In our experiment strategy, each given $Q$, we evaluate various $P$ including $2, 4, 8, 16, 32, 64$, and select the one that achieves the best result on emotion recognition as the appropriate $P$ for given $Q$. 
	The results of the different $Q$ on emotion recognition %with the appropriate number of sampled video clips per minibatch ($P$) for it
 are illustrated in Fig.11.  The appropriate $P$ and number of epochs of pre-training, and corresponding pre-training accuracy of the different $Q$ are reported in Table \uppercase\expandafter{\romannumeral4} .
\par

\par

\par
	
\begin{table}%[b]
\fontsize{9}{10}\selectfont
\centering
\caption{ Illustration of the appropriate combination of hyper parameters of $Q$ and $P$ in the hyper parameter analysis on DEAP and SEED.
%Illustration of the appropriate number of epochs of pre-training for various $Q$ (number of samples per group) and corresponding accuracy of per-traing task. 
%$Epoch_{pre}$ represent the best number of epochs of pre-training, %its corresponding model achieve the appropriate performance on emotion classification after fine-tuning. 
%$acc_{pre}$ represent the accuracy of pre-training task.
}
% of various number of samples per group ($Q$) on downstream performance on DEAP and SEED dataset. Proportion represents the proportion of labeled data in training set. The result is average of five times of downstream training. }
\begin{threeparttable}
 \begin{tabular}{llll|llll}  
\toprule   
 % & & (1)&(2)&(3)&(4)&(5)&(6) \\ 
   \multicolumn{4}{c|}{DEAP}    & \multicolumn{4}{c}{SEED} \\ 
%\cline{1-8}
\midrule   
Q &P &$Epoch_{pre}$&$acc_{pre}$& Q &P &$Epoch_{pre}$&$acc_{pre}$ \\
% & &Y&Y&Y&Y&N&\\
\midrule   
 %Baseline & &  &\textbf{100.00} &00.00&00.00&00.00\\
1&16& 440&70.56&1& 8&1992 & 71.58   \\
 2 & 8  &2800&91.11&2& 16 &3288& 80.68    \\
 3 & 8 & 3600 &87.50& 3&32&1296& 66.24  \\
 4& 4  &800&93.06 &4& 32&744& 69.82 \\
 8 & 4 &  475&96.94&7& 32&548&72.45     \\
 16& 4 &450 &97.08& & &&  \\
%\cline{1-8}
 % \multicolumn{4}{c|}{Fully-supervised}  &33.16& 50.15&79.69&87.24  \\
\midrule   
 %SeqCLR[]&&& 00.00& 00.00&00.00&00.00\\
%  \bottomrule  
\end{tabular}
% 这个地方是引入注释或者标注
\begin{tablenotes}
\item$Epoch_{pre}$ represent the appropriate number of epochs of pre-training, $acc_{pre}$ represent the accuracy of pre-training task, $Q$ represent number of samples per group, $P$ represent number of sampled video clips per iteration.  %[1] A represent B
%\item%[2] B represent C
\end{tablenotes}
\end{threeparttable}
\label{tabXXXXXXX}
\end{table}
\begin{figure}%[H]
\centering
\includegraphics[width=0.47\textwidth,height=0.28\textwidth]{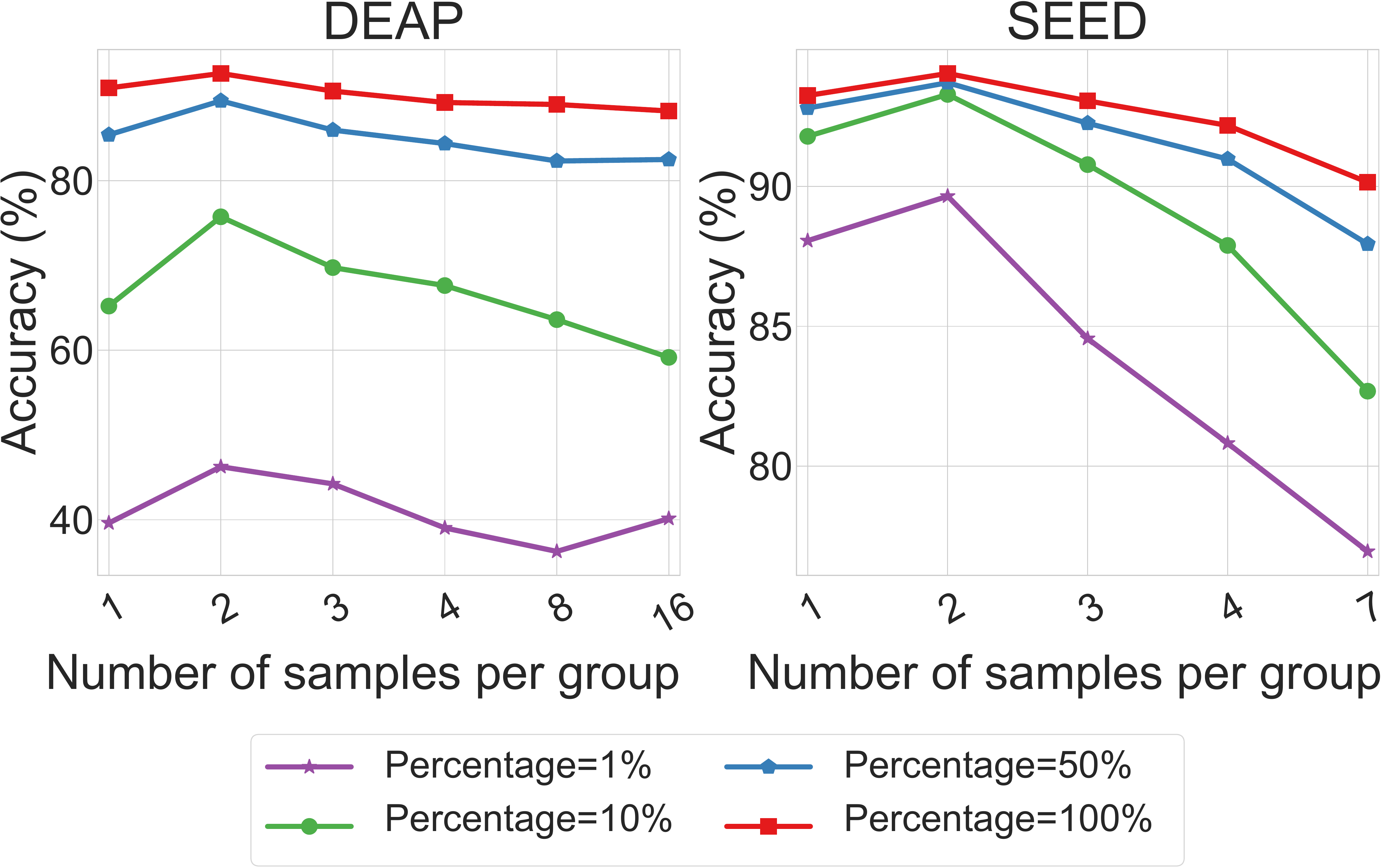}
    \caption{
The group size effect on accuracy of DEAP (left) and SEED (right). The x-axis represents the number of samples ($Q$). The four lines with colors show the various percentage of labeled samples used for fine-tuning in the training set.
%Emotion classification results of various $Q$ (number of samples per group) on DEAP and SEED dataset. $Percentage$ represents the percentage of labeled samples used for supervised fine-tuning in the total training set.
} %The result is average of five times of downstream training.}
\end{figure}
%%%%%%%%%%%%%%%%%%%%%%%%%%%%%%%%%%%%%%%%%%%%%%%%%%%%%第二稿的hyper parameters分析
\par
	On the DEAP, When $Q=2$ and $P=4$, the SGMC achieves the best performance. 
On the SEED, when $Q=2$ and $P=16$, the SGMC achieves the best performance.\par
Further, it can be observed that an opposite law exists in the DEAP and SEED datasets. When given a larger $Q$, the appropriate $P$ on the DEAP tends to be smaller, and on the SEED tends to be larger. The possible reason is the difference in labeling between the two datasets. On the SEED, the emotional labels are labeled by the experiment designer, which is determined by the emotional attribute of the video stimuli. On the DEAP dataset, emotional labels are labeled by the rating of the subjects. Such labeling is more related to the personalized differences of the subject than to the SEED.  
And because the larger $P$, the more difficult the contrastive learning is. At the time the model is more encouraged to focus on extracting stimuli-related features and ignore the personalized features that are irrelevant stimuli. So the larger $P$ lead to better results on the SEED and hinders better results on the DEAP. This indicates that a smaller $P$ should be considered first to use when the data was labeled by the subject, and a larger $P$ should be considered first to use when the data was labeled by the emotional attributes of the stimuli.
\par
Furthermore, it can be found that generally the greater the $Q$ (when $P$ are constant), the greater the accuracy of pre-training. The possible reason is that the greater group sample contains the more comprehensive group-level stimuli-related features to alleviate the interference of random distractions, fatigue, and individual differences. However, good accuracy in pre-training is not always beneficial to emotion recognition. Too smaller $Q$ leads to lower accuracy of pre-training, which hinders the learning of meaningful representation. Too larger $Q$ leads the model to focus on the aggregation of group-level stimuli-related features and leads the base encoder to ignore learning some emotion-related features to hinder better emotion recognition. So it is critical to select an appropriate $Q$ for constructing the group-sample-based contrastive learning.
%%%%%%%%%%%%%%%%%%%%%%%%%%%%%%%%%%%%%第二稿的hyper parameters分析%%%%%%%%%%%%%%%%%%%%%%%%%%%%%%%%%
\noindent
\subsection{Archtechture Design Analysis}
	In this section, we validate our designed choices by control and ablation experiments. We first verify the rationality of the symmetric function we choose. Furthermore, we evaluate the rationality of the strategy of constructing the group sample, utilizing Meiosis augmentation, and constructing the positive-negative pairs. 
\\
(1) Comparison with Various Symmetric Function
\par
The SGMC selects the symmetric function  MaxPool1D to construct the group projector. To verify its rationality, we compare MaxPool1D with a common similar AvgPool1D and an additional opposite MinPool1D which is implemented by taking the minimum value in each dimension of upgraded representations.
 Illustrate in Fig.4. and Table \uppercase\expandafter{\romannumeral4} MaxPool1D is significantly better than others. The possible reason is that MaxPool1D is more beneficial for model selecting emotion-related features to extract from upgraded feature representations. Although MinPool1D also has a selection ability, the features it selects are more detrimental to improving learning emotion-related representation.
 This verifies the rationality of using MaxPool1D to aggregating group features for contrastive learning.
%This validate MaxPool is suitable for aggregating group features for contrastive learning. \\
\begin{table}%[H]
\fontsize{9}{10}\selectfont
\centering
\caption{Illustration of appropriate number of epochs of the pre-training on DEAP and SEED when comparing the symmetric functions. } 
\setlength{\tabcolsep}{3.0mm}{
\begin{threeparttable}
 \begin{tabular}{lll}
\toprule   
\multirow{2}{*}{Symmetric function}&\multicolumn{2}{c}{$Epoch_{pre}$}\\
\cmidrule(lr){2-3} 
 &\multicolumn{1}{c}{DEAP}&\multicolumn{1}{c}{SEED}\\
\midrule   
 	  MinPool1D& 2440  & 1480    \\
 	 AvePool1D& 1720  & 2472  \\
 	  MaxPool1D& 2880 & 3288   \\
%	 LSTM&Overfitting & Overfitting\\
\midrule    
\end{tabular}
%\begin{tablenotes}
%\item$Epoch_{pre}$ represent the appropriate number of epochs of pre-training .
%\end{tablenotes}
\end{threeparttable}
\label{tabXXXXXXXXXX}}
\end{table}
\begin{figure}%[H]
\centering
\includegraphics[width=0.485\textwidth,height=0.28\textwidth]{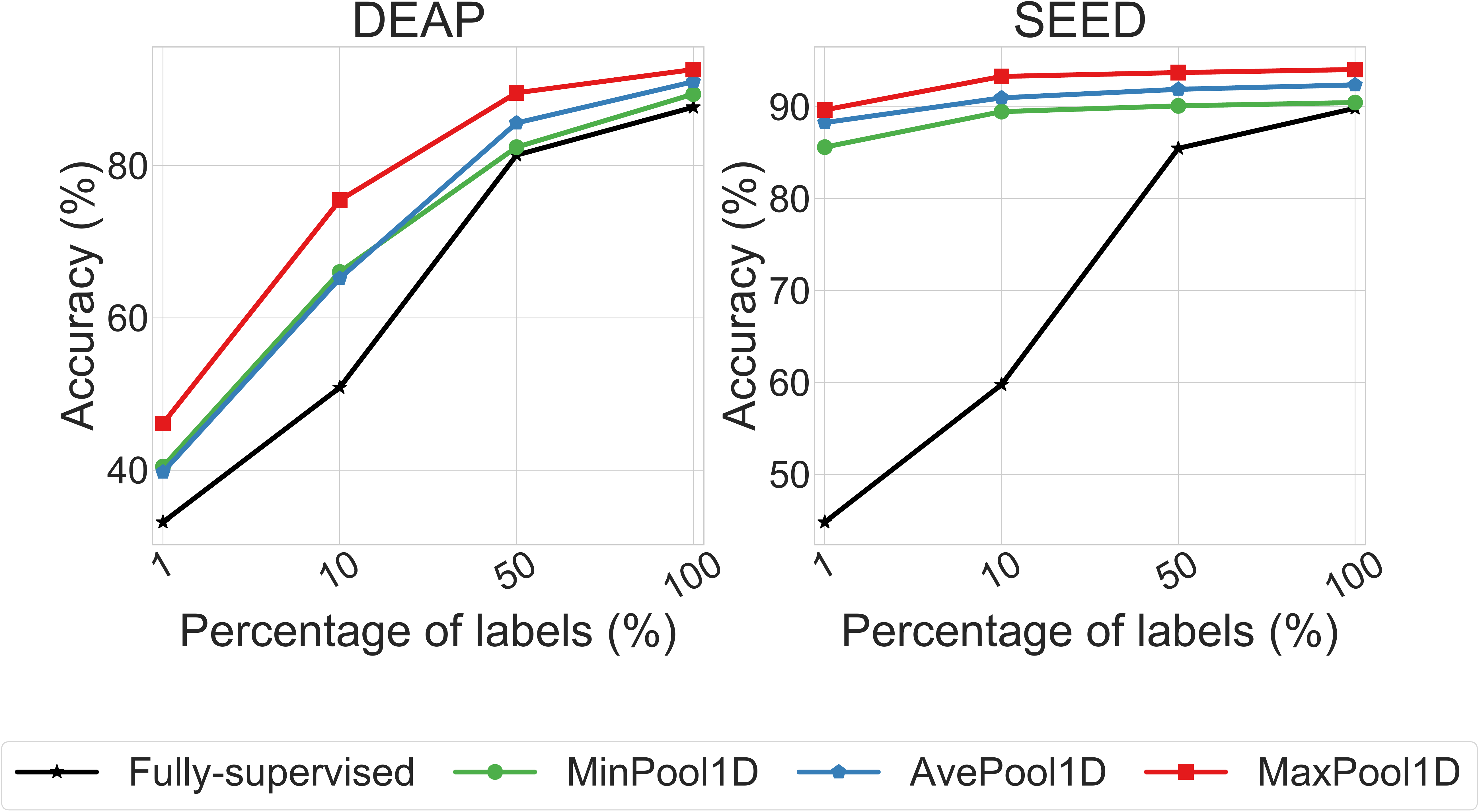}
    \caption{
Emotion classification accuracy based on fully-supervised model and SGMC fine-tuned model with various symmetric functions on DEAP (left) and SEED (right). The x-axis represents the percentages of labeled sample for fine-tuning/supervised training in the training set.
%Emotion classification result of the various symmetric functions. $Percentage$ $of$ $labels$ represents the percentage of labeled samples used for supervised fine-tuning in the total training set.
}
\end{figure}
\\
\noindent (2) Ablation Study
\par
To investigate the rationality of some novel designs of the architecture, we conduct an ablation study for these three components:  group sample, Meiosis data augmentation, and stimuli consistency. 
%When nothing is removed, the detail is consistent with Table \uppercase\expandafter{\romannumeral1}. 
We can get the new version by removing one or two components, and the evaluation strategy is consistent with the basic configuration. When the group sample is ablated, we use individual samples for contrastive learning (just let $Q=1$). When Meiosis data augmentation is ablated, for augmenting the group/individual samples we skip the crossover process and go directly into the separation process after completing individual pairing. After removing the stimuli consistency, we change the way of constructing the positive pair with samples sharing the same stimuli. Instead, the sampler is required randomly sample EEG signals with any stimuli to form the sample group for augmenting and constructing pairs. 
%There are three new versions concluding Non-group Non-aligned and Non-augment by removing these components at one time. 
\par

\begin{table}%[H]
\fontsize{9}{10}\selectfont
\centering
\caption{ The components of the five new versions and the complete SGMC, and the appropriate number of epochs for pre-training with each version on DEAP and SEED. %to achieve the appropriate performance of downstream emotion classification. }%Result of various ablation methods and substitution methods on downstream performance in four proportions of labeled data in the training set. The result is an average of five times downstream training. 
}
% 在tabular 上面引入threeparttable环境
\setlength{\tabcolsep}{1.0mm}{
\begin{threeparttable}
 \begin{tabular}{l|lll|lllllllllllll}
\toprule   
\multirow{2}{*}{Method}&\multirow{2}{*}{Group}&\multirow{2}{*}{Augment}&\multirow{2}{*}{Consistent}&\multicolumn{2}{c}{$Epoch_{pre}$}\\
%\toprule   
 % & & (1)&(2)&(3)&(4)&(5)&(6) \\ 
%\cline{5-6}
\cmidrule(lr){5-6} 
&&&&\multicolumn{1}{|c}{DEAP}&\multicolumn{1}{c}{SEED}\\
%\cline{1-4}
%\midrule   
%\multicolumn{3}{c|}{Percentage of labels}&  \\
%\cline{1-3}
% & &Y&Y&Y&Y&N&\\
\midrule   
 %Baseline & &  &\textbf{100.00} &00.00&00.00&00.00\\
 	  Non-group&\XSolidBrush &Crossover& \Checkmark & 440  & 1848    \\
 	 Non-augment&\Checkmark & No augment & \Checkmark  & 800  & 2280  \\
 	  Mixup-augment&\Checkmark &  Mixup&\Checkmark & 275 & 1304   \\
 	 Non-consistent&\Checkmark &Crossover& \XSolidBrush& 60 & 2752*   \\
 	  Consistent-only&\XSolidBrush &No augment& \Checkmark & 1740  & 2368  \\
 	  Proposed&\Checkmark&  Crossover&\Checkmark &  2800&   3288 \\
%\cline{1-13}
	%\multicolumn{3}{c|}{Fully-supervised}&-&\small{33.16}& 50.85 &81.39 & 87.68&-&44.81& 59.77&85.47&89.83\\
\midrule   
 %SeqCLR[]&&& 00.00& 00.00&00.00&00.00\\
%  \bottomrule  
\end{tabular}
% 这个地方是引入注释或者标注
\begin{tablenotes}
\item * Non-consistent leads to worse performance of emotion recognition than fully-supervised on the SEED dataset, so we adopt the result obtained when the loss function of pre-training converges. 
\item $Epoch_{pre}$  represents the appropriate number of epochs of pre-training.
\item No augmet represent ablating crossover, Crossover represent adopt crossover to data augment ,and Mixup represent adopt Mixup to substitute crossover 
\end{tablenotes}
\end{threeparttable}
\label{tabXXXXXXXXXX}}
\end{table}
\begin{figure}%[H]
\centering
\includegraphics[width=0.485\textwidth,height=0.28\textwidth]{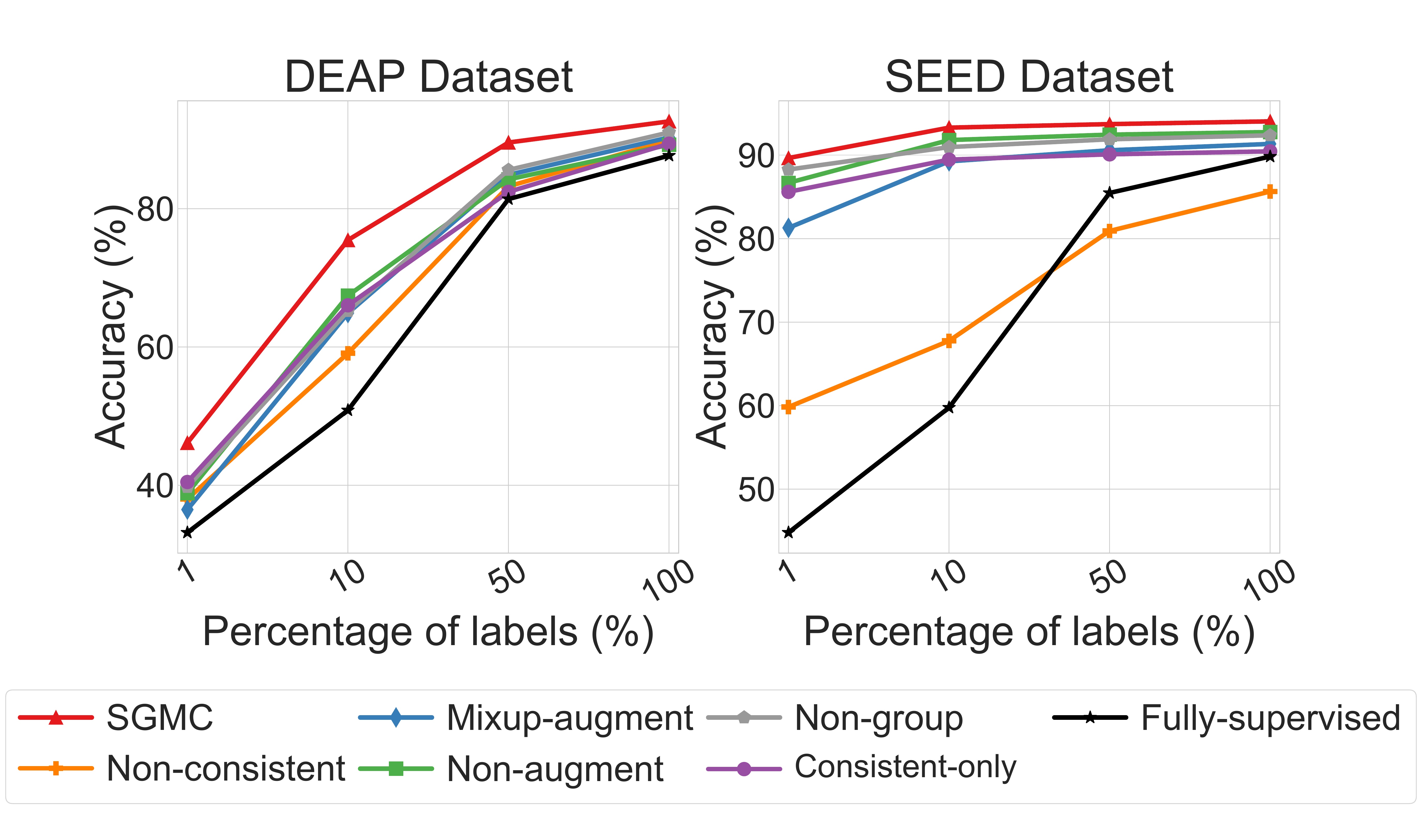}
    \caption{Emotion classification accuracy based on the five new versions fully-supervised, and the complete SGMC on DEAP (left) and SEED (right). The x-axis represents the percentages of labeled sample for fine-tuning/supervised training in the training set. %$Percentage$ $of$ $labels$ represents the percentage of labeled samples which used for supervised fine-tuning in the total training set.
}
\end{figure}
The results of the four-category classification on DEAP and three-category classification on SEED are reported in Fig.11. The detail of ablation and the number of epochs of pre-training are reported in  Table \uppercase\expandafter{\romannumeral6}.
\par
To verify the effectiveness of group sample on the SGMC, we design a version Non-group by removing the group sample.
 It can be observed that the emotion recognition performance significantly declined on DEAP and SEED by more than 1.5\%. This reflects group sample is important to alleviate the obstacles of contrastive learning for the SGMC framework.
%the group sample is beneficial in achieving better emotion recognition results.
To verify the effectiveness of Meiosis augmentation, we design a version Non-augment by removing Meiosis augmentation. It can be observed that on DEAP the accuracy decreases significantly more than 3\%, and on SEED decreases by more than 1.2\%. It verifies the critical role of Meiosis data augmentation to improve emotion recognition in the SGMC.
To verify the superiority of Meiosis utilizing the stimuli alignment in the group sample, we design a competitive version Mixup-augment. For the mutual augmentation of two samples, we can naturally think of Mixup \cite{b37} data augmentation, which can mix two samples and generate two samples. We construct the Mixup-augment version by substituting the crossover of Meiosis with Mixup. The result shows the Meiosis-based SGMC significantly exceeds Mixup-augment by more than 2.3\% on the DEAP and 2.6\% on SEED. This shows the effectiveness of designing the Meiosis data augmentation by mimicking the physiological mechanisms of meiosis.
To verify the importance of constructing the positive-negative pair based on the consistent stimuli, we design a version Non-aligned by removing stimuli consistent. The results significantly decrease by more than 2.7\% on DEAP, even lower than the fully-supervised baseline on SEED. It reflects that stimuli consistent is critical to guiding learning meaningful stimuli-related feature representation by constructing instructive positive-negative pairs.
Further, to investigate the utilizability of potential stimuli consistency, we perform a version Consistent-only by removing the group sample and Meiosis augment, and keeping stimuli consistent only for contrastive learning. The result exceeds the fully-supervised baseline by more than 1.7 \% on DEAP and by more than 0.6\% on SEED, which indicates that consistency of stimuli are exploitable but hindered.

\section*{Conclusion and Future Work}
	In this work, a self-supervised Group Meiosis Contrastive learning (SGMC) framework is designed to %utilize the potential stimuli label to 
improve emotion recognition. %guide the construction of positive-negative pairs for contrastive learning.
%In the proposed framework, Meiosis data augmentation is designed to augment EEG group samples with ensuring unchanged stimuli features. 
In the proposed framework, Meiosis data augmentation is introduced to augment EEG group samples without changing stimuli features.
% And the model consisting of a base encoder and a group projector aims at extracting group-level feature representations.
A base encoder and a group projector are designed in the model to extract group-level feature representations.
 % The potential stimuli labels were utilized to guide the construction of positive-negative pairs for contrastive learning.
With the consistency of stimuli, contrastive learning is designed to learn stimuli-related feature representation.
%Contrastive learning performs under the guidance of potential stimuli labels.
 %In the proposed framework, we developed the data augmentation that can augment the group sample to generate two group samples for constructing positive pairs.
%We designed the group projector which can extract group-level feature representations for calculating the similarity between group samples.  % and take the advantage of the stimuli alignment among sample groups the to augment group sample for contrastive learning. 
\par
%1.我们的框架采用潜在刺激标签来指导对比学习实现了情感识别精度的提升，尤其是在有限标签下框架能很大程度代替人工标签提升精度。2.通过SGMC，模型可以学习到提取视频级的刺激从而提升情感特征表示的能力。3.
	%The experiments demonstrated several conclusions:
The proposed framework achieves state-of-the-art emotion recognition results on the DEAP, and also reaches competitive performance on the SEED datasets. %Compared with the fully-supervised baseline, the SGMC significantly improves emotion recognition to greater classification results, especially when using few artificial labels. 
Compared to the fully-supervised baseline, the SGMC improves emotion recognition significantly, especially when there are limited labels.
In addition, the results of feature visualization suggest that the model might have learned the video-level feature representations, and improves the performance of the model.
%And the feature visualization suggests the possible reason for performance improvement. Through enabling the model to learn video-level feature representations to improve the emotion-related representation ability.
 %Moreover, Meiosis data augmentation shows an effectively genetics-inspired significance in taking advantage of stimuli alignment to generate meaningful group pairs.%positive-negative pairs.
%删Moreover, Meiosis data augmentation shows an effectively genetics-inspired significance in the ablation study. 
%To explore the effectiveness of group samples, hyper parametric analysis further verifies that the group samples be beneficial to emotion recognition. %in certain group sizes.
%based on the guiding of stimuli labels, 
%The application of group-based samples has been revealed to be conducive to emotion recognition in certain combinations of group size and the number of groups. %We further verified 
%the application of group samples has been revealed to be conducive to emotion recognition in a certain group size.
%Further, the experiment demonstrates that the group samples be beneficial to emotion recognition.% in certain group sizes.
The hyper parametric analysis further demonstrates the role of group samples during emotion recognition.
Finally, the rationality of the framework design including the selection of symmetric functions, the construction of the positive-negative pairs, and Meiosis data augmentation are verified.
%Finally, it is verified that rationality existing in the organic combination of group-based samples, Meiosis, and based stimuli-label-guided contrastive learning. %And based on the guiding of stimuli labels, we verified that the group sample is more efficient than the individual sample to extract the stimuli-related representation for contrastive learning.
%1) the potential stimuli labels can be utilized to guide contrastive learning to achieve good improvement in emotion recognition.
%2) The SGMC could achieve excellent performance on limited labeled sample learning. 
%3) The feature visualization verified the SGMC enables model learning video-level feature representations to improve emotion recognition.
%4) The group sample is more efficient than the individual sample to extract the stimuli-related representation for contrastive learning.
%5) The Meiosis data augmentation could take advantage of stimuli alignment among a group of samples to generate meaningful positive pairs.
%have effectiveness in augmenting group samples for contrastive learning.
%6) The feature visualization shows that the model learned distinguishable video-related feature representations through the SGMC. 
\par
%Although the SGMC enable effectively makes up for the scarcity of labels, a problem of high consumption in the calculation is existing.
In the future, we will continue to develop such kinds of group-sample-based SSL frameworks while with low calculation costs.

\end{document}